\documentclass[a4 paper,11 pt]{article}
\pdfoutput=1 

\usepackage{jheppub} 

\usepackage[T1]{fontenc}
\usepackage{amsmath}
\usepackage{physics}
\usepackage{cancel}
\usepackage{caption}
\usepackage{subcaption,longtable,stmaryrd}
\usepackage{comment}

\usepackage{subcaption}
\usepackage{float}

\DeclareSymbolFont{usualmathcal}{OMS}{cmsy}{m}{n}
\DeclareSymbolFontAlphabet{\mathcal}{usualmathcal}

\usepackage{bigints}
\usepackage{multicol}
\usepackage{blindtext}
\usepackage{tikz}
\usepackage{enumitem}
\usepackage{amsmath, amssymb, graphics, setspace}
\usepackage{float}
\newcommand{\mathsym}[1]{{}}

\newcommand{\unicode}[1]{{}}
\usepackage[mathscr]{euscript}
\DeclareSymbolFont{rsfs}{U}{rsfs}{m}{n}
\DeclareSymbolFontAlphabet{\mathscrsfs}{rsfs}
\DeclareSymbolFont{rmlargesymbols}{OMX}{mdbch}{m}{n}
\DeclareMathSymbol{\rmintop}{\mathop}{rmlargesymbols}{82}
\DeclareMathSymbol{\rmointop}{\mathop}{rmlargesymbols}{72}
\usepackage[customcolors,shade]{hf-tikz}

\usepackage{graphicx}

\definecolor{darkbrown}{rgb}{0.787, 0.26, 0.187}
\definecolor{alizarin}{rgb}{0.82, 0.1, 0.26}

\title{\bf{Dynamics of monitored SSH Model in Krylov Space: From Complexity to Quantum Fisher Information}}

\author[]{Nilachal Chakrabarti,}
\author[]{Neha Nirbhan,}
\author[]{and Arpan Bhattacharyya}

\affiliation[]{Department of Physics, Indian Institute of Technology Gandhinagar, Gujarat 382055, India.}

\emailAdd{nilachalchakrabarti@iitgn.ac.in}
\emailAdd{neha.nirbhan1207@gmail.com}
\emailAdd{abhattacharyya@iitgn.ac.in}
\abstract{In this paper, we investigate the dynamics of a non-Hermitian Su-Schrieffer–Heeger model that arises out of the no-click limit of a monitored SSH model in the Krylov space. We find that the saturation timescale of the complexity associated with the spread of the state in the Krylov subspace increases with the measurement rate, and late time behaviour differs across the  $\mathrm{PT}$ symmetry transition point. Furthermore, extending the notion of this complexity for subsystems in Krylov space, we find that the scaling of its late time value with subsystem size shows a discontinuous jump across the $\mathrm{PT}$ transition point, indicating that it can be used as a suitable order parameter for such transition but not for the measurement-induced transition. \textcolor{black}{Finally, we show that a generalized measure in the Krylov subspace, which contains information about the correlation landscape, such as  Quantum Fisher information, which also possesses some structural similarity with the complexity functional, can be a promising probe of the measurement-induced phase.}}

\makeatletter

\begin{document}

\maketitle

\section{Introduction}
The concept of complexity is of current interest in theoretical physics, which has been used in several areas of physics, ranging from simple quantum mechanical models to complex quantum field theoretical and holographic models to study (mainly) the dynamical behaviour of chaotic systems. There are various notions of complexity. One such notion is circuit complexity. It was defined by Nielsen et al. as a measure of the difficulty in reaching a target state from a reference state and tells us that finding the optimal quantum circuit is equivalent to finding the shortest path between two points in a certain curved geometry \cite{NL1,NL2,NL3}. Over the past several years, it has been used and explored for various quantum mechanical and field-theoretic models \cite{Jefferson,Chapman:2017rqy,Bhattacharyya:2018wym,Caputa:2017yrh,Ali:2018fcz,Bhattacharyya:2018bbv,Hackl:2018ptj,Khan:2018rzm,Camargo:2018eof,Ali:2018aon,Caputa:2018kdj,Guo:2018kzl,Bhattacharyya:2019kvj,Erdmenger:2020sup,Ali:2019zcj,Bhattacharyya:2019txx,Caceres:2019pgf,Bhattacharyya:2020art,Liu_2020,Bhattacharyya:2020rpy,Bhattacharyya:2020kgu,Chen:2020nlj,Czech:2017ryf,Chapman:2018hou,Couch:2021wsm,Chagnet:2021uvi,Koch:2021tvp,Bhattacharyya:2022ren,Bhattacharyya:2023sjr,Bhattacharyya:2022rhm,Bhattacharyya:2021fii,Bhattacharyya:2020iic,Craps:2023rur,Jaiswal:2021tnt,Bhattacharya:2022wlp,Bhattacharyya:2024rzz,Chowdhury:2023iwg}\footnote{This list is by no means exhaustive. Interested readers are referred to these reviews and thesis \cite{Chapman:2021jbh, Bhattacharyya:2021cwf,Katoch:2023etn}, and references therein for more details.}. Also, it has found significant application in the context of AdS/CFT correspondence, where it can be related to the growth of the Einstein-Rosen bridge in the dual bulk side \cite{Susskind:2014moa,Brown:2015bva,Stanford:2014jda,Carmi:2016wjl}.\par  

However, a new idea of complexity named Krylov complexity has recently been adapted \cite{PhysRevX.9.041017, Barbon:2019wsy, Rabinovici:2020ryf}. From the initial lessons of quantum mechanics, we know that given an initial local operator $\mathcal{O}(0)$ at time $t=0$, it will typically evolve to some complex operator $\mathcal{O}(t)$ under the time-evolution generated by the  Hamiltonian $H.$ In Heisenberg picture,the operator $\mathcal{O}$(t) is given by, $\mathcal{O}(t)= e^{-iHt}\mathcal{O}(0) e^{iHt}$. Using the Baker-Campbell-Hausdorff formula, it can be shown that the operator $\mathcal{O}(t)$ can be written in terms of nested commutators with increasing complexity. The growth of the operator describes how an initially localized operator spreads to the entire system. This new notion of complexity has been used to study thermalization and to investigate the integrable to chaotic transition in quantum many-body systems \cite{PhysRevX.9.041017, Barbon:2019wsy, Rabinovici:2020ryf}\,. 

A ``Universal Operator growth'' hypothesis defines how to distinguish between a chaotic and integrable system \cite{PhysRevX.9.041017} where the authors have studied the operator growth under a generic,non-integrable Hamiltonian. The hypothesis is based on a recursive technique termed ``Lanczos Algorithms''. Using the Lanczos Algorithm, one gets orthogonal basis vectors $\ket{\mathcal{O}_{n}}$ and a set of Lanczos coefficients $\{b_{n}\}$. The operator $\mathcal{O}(t)$ then can be written in terms of Krylov basis $\ket{\mathcal{O}_{n}}$ and operator wave function $\phi_{n}(t)$ can be shown to obey a discrete Schrodinger equation. This hypothesis argued that $b_{n}$ will grow at most linearly with $n$ for a chaotic model. For integrable models, $b_{n}\sim \alpha n^{\delta}$,where $0<\delta <1$. This implies that the Krylov complexity grows exponentially with time for chaotic models and power-like for integrable models \cite{PhysRevX.9.041017}. This was studied mainly for Random Matrix models, different spin chain models (fermionic), including the celebrated $\textrm{SYK}$ model and certain bosonic lattice models like Bose-Hubbard Models as well for certain quantum mechanical and simple quantum field theory models. Interested readers can go through the list of references \cite{Barbon:2019wsy,Jian:2020qpp,DymarskyPRB2020,PhysRevLett.124.206803,Yates2020,Rabinovici:2020ryf,Rabinovici:2021qqt,Yates:2021lrt,Yates:2021asz,Dymarsky:2021bjq,Noh2021,Trigueros:2021rwj,Liu:2022god,Fan_2022,Kar_2022,Caputa:2021sib,PhysRevE.106.014152,Bhattacharjee_2022a,Adhikari:2022whf,https://doi.org/10.48550/arxiv.2205.12815,Bhattacharya:2022gbz,H_rnedal_2022,Bhattacharjee:2022lzy,Rabinovici:2022beu,Alishahiha:2022anw, Avdoshkin:2022xuw,Camargo:2022rnt,Bhattacharjee:2023uwx,Iizuka:2023pov, Iizuka:2023fba, Rabinovici:2023yex,Zhang:2023wtr,Hashimoto:2023swv,Erdmenger:2023wjg,Bhattacharyya:2023dhp,Alishahiha:2024rwm,Menzler:2024ifs,Loc:2024oen}\,.This list is by no means exhaustive. Interested readers are referred to these reviews and thesis \cite{Nandy:2024htc, Sanchez-Garrido:2024pcy}, and references therein for more details.
\par
Recently, this idea of complexity in Krylov space has been extended for states. In \cite{Balasubramanian:2022tpr}, the authors have shown that the spread complexity quantifies the spread of an initial quantum state under a Hamiltonian in the Krylov basis and this has given new directions into the dynamics of scrambling and phase transitions in chaotic systems \cite{Balasubramanian:2022tpr,Bhattacharjee:2022qjw, Nandy:2023brt,Banerjee:2022ime,PhysRevD.108.025013,Nizami:2023dkf,PhysRevB.108.104311,PhysRevB.109.014312,PhysRevB.109.104303,PhysRevLett.132.160402,Caputa:2024vrn,PhysRevB.110.064318,Nizami:2024ltk,Balasubramanian:2023kwd,Baggioli:2024wbz,Fu:2024fdm}.
Furthermore, the spread complexity has also been used as a possible indicator for detecting certain quantum phase transitions \cite{Caputa2022PRB,Anegawa:2024wov}, although much investigation is still needed to concretely establish spread complexity as a reliable probe for quantum phase transitions. In recent times, investigation of the spread complexity was extended to certain open systems \cite{Bhattacharya:2023zqt,Bhattacharyya:2023grv,Carolan:2024wov} as well to certain simple non-Hermitian  models \cite{Bhattacharya:2023yec,Bhattacharya:2024hto,Sahu:2024urf} where a generalized version of Lanczos algorithm was required to compute the spread complexity. 
\par
Monitored quantum spin chain models are spin chain models that undergo unitary evolution by a Hamiltonian with some local measurements on the chain, which introduce a non-unitary evolution in the dynamics. Recently, entanglement transition(s)  have been getting attraction for these monitored spin chain models as well as for certain random circuits \cite{skinner2019measurementinduced,li2018quantum,li2019measurementdriven,chan2019unitaryprojective,boorman2022diagnostics,Biella2021manybodyquantumzeno,szyniszewski2020universality,barratt2022transitions,zabalo2022infinite,barratt2022field,lunt2020measurement,turkeshi2021measurementinduced,zabalo2022operator,iaconis2021multifractality,Sierant2022dissipativefloquet,bao2020theory,choi2020quantum,szyniszewski2019entanglement,block2022measurementinduced,jian2020measurementinduced,agrawal2021entanglement,gullans2020scalable,sharma2022measurementinduced,zabalo2020critical,vasseur2019entanglement,li2021conformal,turkeshi2020measurementinduced,lunt2021measurementinduced,sierant2022universal,gullans2020dynamical,PhysRevB.107.L220201} \footnote{This list is by no means exhaustive. Interested readers are referred to references and citations of these papers.}. The dynamics of these monitored spin-chain models are described by the stochastic Schrodinger equation, in which the effect of measurement is introduced through a stochastic variable. In the ``no-click'' limit \cite{Wiseman2009,carmichael2009open, gardiner2004quantum,daley2014quantum,Jacobs2014,PhysRevB.105.205125}, the stochastic variable is set to zero. Then, the dynamics of the monitored models can be described by an effective non-Hermitian Hamiltonian. The parameter in front of this non-Hermitian part can be interpreted as the measurement rate. Now the entanglement transition can be characterized by studying the scaling law of entanglement entropy (\textrm{EE}) of the subsystem. Depending on the measurement rate,  \textrm{EE}  grows as the volume or area (or logarithmically at the critical point) of the subsystem. The transition between one scaling law and another is known as the entanglement transition. This kind of transition can be found in monitored quantum circuits \cite{fisher2022quantum,lunt2021quantum,potter2021entanglement} and monitored fermionic chains \cite{alberton2021entanglement,buchhold2021effective,muller2022measurementinduced,turkeshi2021entanglement,turkeshi2021measurementinduced2,Kalsi_2022,kells2021topological,fleckenstein2022nonhermitian,Zhang2022universal,turk} even in the ``no-click'' limit \cite{ashida2018full,bacsi2021dynamics,dora2021correlations, gopalakrishnan2021entanglement,10.21468/SciPostPhys.14.5.138} as a function of measurement rate.  
\par
Motivated by these works, we initiate a study of the dynamics of the non-Hermitian $\textrm{SSH}$ model in the Krylov space. Our model possesses a $\mathrm{PT}$ symmetry phase (where the eigenspectrum of the system becomes purely real) and also displays an entanglement transition in the  $\mathrm{PT}$ broken phase \cite{10.21468/SciPostPhys.14.5.138}. We use spread complexity and associated measures, e.g., spread entropy, to probe this system to see whether it can be a useful indicator for detecting both of these phase transitions, i.e., $\mathrm{PT}$  and entanglement transition with the goal to demonstrate that indeed the spread complexity functional is a useful tool for detecting phase transition adding to the previous results. Furthermore, given the structural similarity of \textit{Quantum Fisher Information} (QFI) \cite{Shi:2024bpu, PhysRevLett.133.110201} in the Krylov space, with that of the Krylov complexity functional, we explore this quantity in the Krylov space as it contains the information about the correlation landscape of the underlying system. 
\par
The paper is organized as follows. In Sec.~(\ref{section1}), we give a brief overview of our model. We also discuss the change in the eigen-spectrum w.r.t the non-Hermitian parameter and how it is related to the two transitions mentioned earlier. In Sec.~(\ref{section3}), we present the analysis of spread complexity, spread entropy, and entropic complexity in the $\mathrm{PT}$ symmetric and $\mathrm{PT}$ broken regions. Furthermore, we extend the computation of spread complexity for a subsystem in the Krylov space and discuss its late-time scaling (w.r.t the subsystem size) and whether it can detect the phase transitions. In Sec.~(\ref{section-6}),  we will discuss the ``Quantum Fisher Information'' (\textrm{QFI}) in the Krylov space and its relation with Krylov complexity. Furthermore, we will discuss the variation of averaged $\textrm{QFI}$ with the non-Hermitian parameter (measurement rate) and whether it can probe the two transitions. Finally, in Sec.~(\ref{section-7}), we summarize our results once again and point out some future directions. Details of the bi-Lanczos algorithm used in this paper and other useful details are given in Appendix~(\ref{app-A}) and Appendix~(\ref{app-B})\,.
\section{Brief overview of non-Hermitian SSH model}\label{section1}
\begin{figure}
    \centering
    \includegraphics[width=0.5\linewidth]{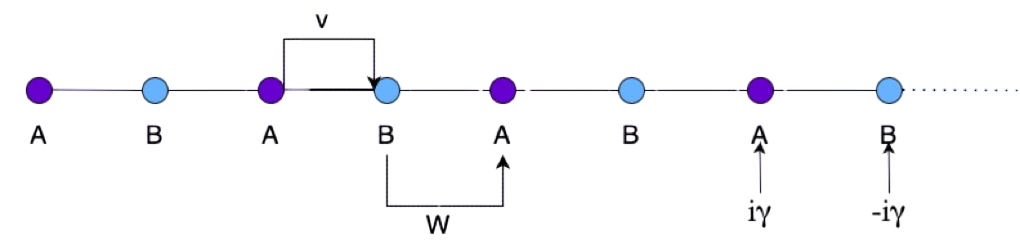}
    \caption{Schematic diagram of non-Hermitian SSH model.}
    \label{fig1}
\end{figure}

In this section, we will briefly review the model of our interest, i.e. the non-Hermitian Su-Schrieffer–Heeger (SSH) Model \cite{10.21468/SciPostPhys.14.5.138,PhysRevB.103.085137,PhysRevB.97.045106,PhysRevB.110.115135,Rottoli_2024}. It is a fermionic chain with two different sublattices: $A$ and $B$ of length $L\,.$ The fermionic annihilation and creation operators are denoted by $c_{\sigma,j}$ and $c^{\dag}_{\sigma,j}$ where $\sigma=\{A,B\}$ and $j=\{1,2,\ldots L\}$. The number operator for the $j^{th}$ site is denoted by $n_{\sigma,j}=c_{\sigma,j} c^{\dag}_{\sigma,j}$. Then the Hamiltonian is given by \cite{10.21468/SciPostPhys.14.5.138}, 
\begin{equation}\label{ham_1}
   H=\sum_{j=1}^{L}\Big(-wc^{\dag}_{A,j}c_{B,j}-vc^{\dag}_{A,j}c_{B,j+1}+\textrm{h.c.}\Big)-i\gamma
   \sum_{j=1}^{L}\Big(n_{A,j}-n_{B,j}\Big)\,.
\end{equation}
Here, $w,v$ denote the intra and inter-cell hopping parameters (refer to Fig.~(\ref{fig1})), respectively, and $\gamma$ (>0) introduces the non-hermiticity in the model. In our subsequent computations, we will represent this Hamiltonian in terms of a $2L \times 2L$ matrix using the single-particle basis and use the open boundary condition to study the dynamics. 
This model possesses an intrinsic $\mathrm{PT}$ symmetry. The action of parity ($\mathrm{P}$) and time-reversal ($\mathrm{T}$) is defined by \cite{PhysRevLett.80.5243},
\begin{equation}
  \mathrm{P}c_{A,j}\mathrm{P}=c_{B,L-j+1},\quad \mathrm {P}c_{B,j}\mathrm{P}=c_{A,L-j+1},\quad \mathrm {T}i\mathrm{T}=-i\,.
\end{equation}
Using these definitions, it follows that the non-Hermitian SSH model remains invariant under the combined operation of  $\mathrm{P}$ and $\mathrm{T}$, i.e.,
 $$ (\mathrm{PT})\,H\,(\mathrm{P T})^{\dagger}=H\,.$$
Moreover, this model  preserves the total particle number,
\begin{equation}
    Q=\sum_{j=1}^{L}\Big(n_{A,j}+n_{B,j}\Big)\,.
\end{equation}
\vspace{0.2cm}\\
It can be shown that the non-Hermitian Hamiltonian can be used to describe the quantum dynamics of the system by continuously monitoring Hermitian Models in the no-click limit where the non-Hermitian parameter is equivalent to the measurement rate \cite{10.21468/SciPostPhys.14.5.138}. Starting with an initial state $\ket{\psi(0)}$, we can solve for $\ket{\psi(t)}$ using a deterministic equation, which is the no-click limit of the stochastic Schr\"{o}dinger equation,
\begin{equation}
    d\ket{\psi(t)}=-iH dt\ket{\psi(t)}-i\frac{dt}{2}\langle H-H^{\dag}\rangle_{t}\ket{\psi(t)}\,.
\end{equation}
Solving this equation, $\ket{\psi(t)}$ can be written as,
\begin{equation}\label{Evolpe eqn}
    \ket{\psi(t)}=\frac{e^{-iHt}\ket{\psi(0)}}{||e^{-iHt}\ket{\psi(0)}||}\,.
\end{equation}
Recently, it has been shown in \cite{Bhattacharya:2024hto} that for certain simple non-Hermitian spin models (e.g. tight-binding model), perhaps one can study the dynamics in the Krylov space and use the spread complexity, i.e. the complexity associated with the spread of state in the Krylov space to probe the transition across the  $\mathrm{PT}$ broken and non-broken phases. However, we still need more studies to establish whether the spread complexity can be used to detect this kind of   $\mathrm{PT}$ phase transition or, rather, any kind of phase transition. Motivated by this, we will now study the quantum dynamics of the non-Hermitian SSH model in the Krylov space. Interestingly, this model not only shows a $\mathrm{PT}$ transition but also gives rise to a measurement-induced entanglement phase in the $\mathrm{PT}$ broken phase where the bipartite entanglement entropy for a subsystem, displays a transition from,  volume to area law (w.r.t to the subsystem size) \cite{10.21468/SciPostPhys.14.5.138}.
One of our motivations is to investigate whether, using spread complexity, we can detect both the $\mathrm{PT}$ transition as well as the entanglement transition. If the answer turns out to be negative, we would like to see what other relevant measure(s) in Krylov space can be used to detect this transition, particularly the entanglement one.  
Before studying the dynamics of this model in the Krylov space, we will briefly discuss the eigen-spectrum of this model to get an insight into these transitions.


\subsection*{Eigenspectrum of non-Hermitian SSH model:}\label{section2}
The eigen-spectrum of the non-Hermitian Hamiltonian is generally complex. The imaginary part of the spectrum represents that the probability amplitude of the wave function will decay after a finite lifetime. The scenario is different when the system has a $\textrm{PT}$ symmetry, which makes the eigen-spectrum real.  Since our model, as mentioned in (\ref{ham_1}), is quadratic in fermionic operators, the eigen-spectrum can be solved exactly. To solve this, we first perform a discrete Fourier transformation,
\begin{equation}
    c_{\sigma,j}=\sum_{k}\frac{1}{\sqrt{L}} e^{ikj}\tilde{c}_{\sigma,k},\hspace{0.3cm} k\in\{-\pi,\pi\},\, \sigma= A,B\,.
\end{equation}
Then the Hamiltonian in (\ref{ham_1}) can be re-casted in the form,
\begin{align}
\begin{split}
 &   H=\sum_{K}\begin{pmatrix}
        \tilde{c}_{A,k}^{\dag} && \tilde{c}_{B,k}^{\dag} 
    \end{pmatrix}\mathcal{H}_{k}
    \begin{pmatrix}
        \tilde{c}_{A,k}^{\dag} \\
        \tilde{c}_{B,k}^{\dag} 
    \end{pmatrix}\,,\\&
    \mathcal{H}_{k}=\begin{pmatrix}
        i\gamma && \eta_{k}\\
        \eta_{k}^{*}&& -i\gamma
    \end{pmatrix} \hspace{0.5cm}\eta_{k}=-w-ve^{-ik}\,.
\end{split}
\end{align}
The energy spectrum depends on the eigenvalues of the single-particle Hamiltonian $\mathcal{H}_{k}$. Diagonalizing the Hamiltonian, we can write the eigenvalues as,
\begin{equation}
    \epsilon_{\pm,k}=\pm \sqrt{|\eta_{k}|^{2}-\gamma^{2}}=\pm\sqrt{(w^{2}+v^{2}-\gamma^{2})+2wv\cos{k}}\,.
\end{equation}
Depending on the values of $w,v$ and $\gamma$, the eigen-spectrum can be real or imaginary, i.e., the system will be in $\mathrm{PT}$ broken or $\mathrm{PT}$ symmetric phase. The critical value for $\mathrm{PT}$ transition is given by $\gamma=|w-v|$ \cite{10.21468/SciPostPhys.14.5.138}. When $(w-v)>\gamma$ and $(w-v)<-\gamma$, the system is in $\mathrm{PT}$-unbroken phase. If $-\gamma<(w-v)<\gamma$, the system will be in $\mathrm{PT}$ broken phase.
\begin{figure}[htb!]
    \centering
    \subcaptionbox{$\gamma=0.5$}[0.45\linewidth]{\includegraphics[width=1.1\linewidth]{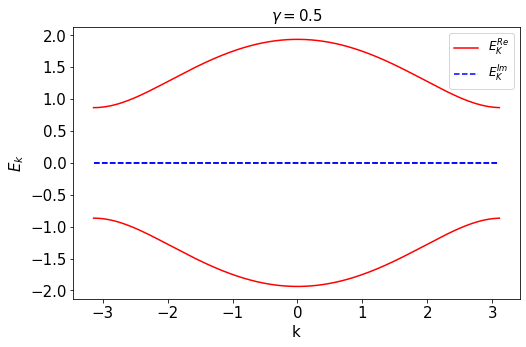}}
    \hspace{0.05\linewidth}
    \subcaptionbox{$\gamma=1.4$}[0.43\linewidth]{\includegraphics[width=1.1\linewidth]{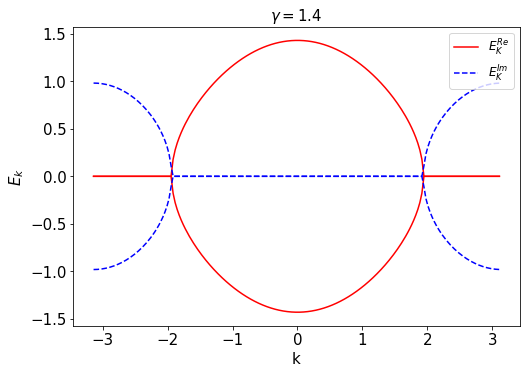}}
    \hspace{0.05\linewidth}
    \subcaptionbox{$\gamma=2.4$}[0.45\linewidth]{\includegraphics[width=1.1\linewidth]{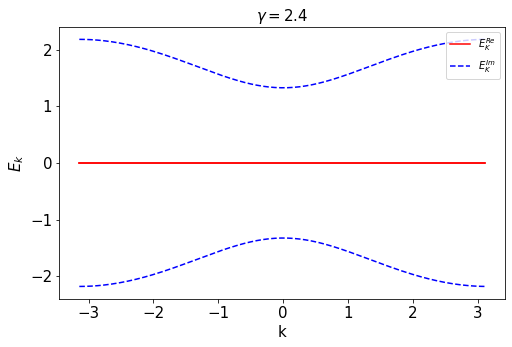}}
    \hspace{0.05\linewidth}
    \caption{The spectrum of the non-Hermitian SSH model for three different values of $\gamma$=0.5\,,1.4\,,2.4 (solid and dashed lines indicate the real and imaginary part of the spectrum, respectively).}
    \label{Fig-1}
\end{figure}
\par
Following \cite{10.21468/SciPostPhys.14.5.138}, we will set the hopping parameters to be $w =1.5,\,v=0.5$ and vary $\gamma$ to see how the real and imaginary part of $\epsilon_{k}$ change. Then the critical value of $\gamma$ for $\mathrm{PT}$ transition is given by $\gamma=1$. As we can see from Fig.~(\ref{Fig-1}), for $\gamma<1$, the spectrum is purely real, and there is a finite gap at $k=\pm \pi$. At the $\mathrm{PT}$ transition point $\gamma=1$, the gap closes at the edge, and two exceptional points occur at $k=\pm \pi$. Beyond the $\mathrm{PT}$ broken region, the spectrum becomes gapless both in real and imaginary parts, and exceptional points occur at $k=\pm k_{\mathrm{EP}}(\gamma)$, whose value can be determined by \cite{10.21468/SciPostPhys.14.5.138},
\begin{equation}
    k_{\mathrm{EP}}(\gamma)=\pm 2\cos^{-1}\sqrt{\frac{\gamma^{2}-(w-v)^{2}}{4wv}}\,.
\end{equation}
Putting $w=1.5$ and $v=0.5$, it can be seen that $k$ will move towards 0 when $\gamma=2$. This hints towards a second transition in the spectrum when all the eigenvalues become purely imaginary and gapped. This precisely corresponds to the entanglement transition as confirmed in \cite{10.21468/SciPostPhys.14.5.138} by explicitly checking the late time scaling of the bipartite entanglement entropy. Now the question is, can we capture the $\mathrm{PT}$ transition and the entanglement transition with the help of spread complexity and other relevant measures? We will try to find the answer to these questions in the next sections.

\section{Spread Complexity for non-Hermitian SSH model}\label{section3}
We now turn our attention to the Krylov Complexity \cite{PhysRevX.9.041017}. There are two notions of Krylov complexity. The first one is associated with the measures of operator growth in the Hilbert space under time evolution. But we will focus on the second one, namely the spread complexity, the complexity associated with the spread of a state in the Krylov basis under time evolution \cite{Balasubramanian:2022tpr}. The key idea behind this is to find the minimal subspace accessible through the time-evolution of a given initial state $|\psi(0)\rangle\,.$ This subspace is known as `\textit{Krylov Space}' and is spanned by a set of orthogonal basis vectors, which can be found using the powerful algorithm known as `\textit{Lanczos algorithm}' \cite{viswanath1994recursion}. However, the \textcolor{black}{standard} Lanczos algorithm is suitable for generating the Krylov basis only when the time evolution is governed by a hermitian Hamiltonian, i.e., when the evolution is unitary. Hamiltonian takes a tri-diagonal form when expressed in terms of the Krylov basis vectors. One needs to modify the Lanczos algorithm when the evolution is governed by a non-Hermitian Hamiltonian, which is the case considered in this paper. 
Instead of the usual Lanczos algorithm, it is more suitable to use the `\textit{bi-Lanczos algorithm}' \cite{bilanczos,Bhattacharya:2023zqt} \footnote{\textcolor{black}{As mentioned in \cite{Bhattacharya:2024hto}, one may still use the complex-symmetric Lanczos Algorithm for a complex-symmetric Hamiltonian. However, keep in mind the possibility of generalizing our studies (especially the one presented in the subsequent \textcolor{black}{Section}~(\ref{section-6})) presented in this paper for more general scenarios where a complex-symmetric Hamiltonian does not necessarily govern the time evolution; we will focus on a more general algorithm, such as the bi-Lanczos algorithm which will be applicable for a wide range of scenarios albeit of the fact that it is computationally more involved.}}. The key idea is to generate two sets of bi-orthonormal vectors $\{|r_{j}|\rangle\}$ and $\{|l_{j}|\rangle\}$ which satisfy $\langle l_i|r_j\rangle=\delta_{ij}\,.$ Now, when one expresses the non-Hermitian Hamiltonian in terms of these bi-orthonormal basis vectors, it takes a tri-diagonal form like the Hermitian case \cite{Bhattacharya:2022gbz}. We sketch the details of the \textit{bi-Lanczos algorithm} in the Appendix~(\ref{app-A})\,.
\par

We will now apply this bi-Lanczos algorithm for our model (\ref{ham_1}). Starting with an initial state $\ket{\psi(0)}=\ket{r_{1}}=\ket{l_{1}}$ which is a product state in our case, we will generate the right and left Krylov basis $\{r_{j}\}$ and $\{l_{j}\}$. Then we can expand the time-evolved wave function $\ket{\psi(t)}$ in terms of $\{|r_{j}\rangle\}$ and $\{|l_{j}\rangle\}$ in the following way,
\begin{equation}
    \sum_{j=1}^{\mathcal{K}}\Psi_{j}^{r}(t)\ket{r_{j}}=\ket{\psi(t)}=\sum_{j=1}^{\mathcal{K}}\Psi_{j}^{l}(t)\ket{l_{j}}
\end{equation}
where $\mathcal{K}$ is the dimension of the Krylov space. The probability amplitude is defined as \cite{Bhattacharya:2022gbz},
\begin{equation}
    P_j(t)=\Big|\tilde{\Psi}_{j}^{r *}(t) \tilde{\Psi}_{j}^{l}(t)\Big|
\end{equation}
where $\tilde{\Psi}_{j}^{p}$ and $\tilde{\Psi}_{j}^{q}$ are the normalized wave function in Krylov Basis. As explained in Appendix~(\ref{app-B}), the normalisation is done dynamically. This ensures the fact that $\sum_{j}^{\mathcal{K}}P_j(t)=1\,.$ 
Then, following \cite{Bhattacharya:2023zqt, Balasubramanian:2022tpr}, we can define the  \textit{spread complexity} and \textit{spread entropy} are defined as,
\begin{align}\label{defination}
\begin{split}
   & \mathcal{C}(t)=\sum_{n}n \Big|\tilde{\Psi}_{n}^{r *}(t) \tilde{\Psi}_{n}^{l}(t)\Big|\,,\\&
    \mathcal{S}(t)=-\sum_{n}\Big(|\tilde{\Psi}_{n}^{r*}(t) \tilde{\Psi}_{n}^{l}(t)|\Big)\ln{\Big[|\tilde{\Psi}_{n}^{r *}(t) \tilde{\Psi}_{n}^{l}(t)|\Big]}\,.
\end{split}
\end{align}
Like the closed system \cite{Balasubramanian:2022tpr}, 
it was shown in \cite{Carolan:2024wov}, that the $\mathcal{C}(t)$ gets minimized when the $|\psi(t)\rangle$ is expanded in terms of the bi-orthonormal Krylov basis vectors. Also, one can define a \textit{entropic complexity} \cite{Balasubramanian:2022tpr},
\begin{equation} \label{definition1}
    \mathcal{C}_S(t)= e^{\mathcal{S}(t)}
\end{equation}
where $\mathcal{S}(t)$ is defined in (\ref{defination}).
This quantity refers to the minimum Hilbert space dimension required to store the information about the 
the probability distribution of Krylov basis weights \cite{Balasubramanian:2022tpr}. \textcolor{black}{To define the strength of the localization of the wave function in the Krylov space, one can define a quantity termed as ``Krylov Inverse Participation Ratio" (\textrm{KIPR}) in the following way\,,} 
\begin{equation}\label{KIPR eqn}
  \textcolor{black}{\textrm{KIPR}_{\textrm{R}}(t)= \sum_{j=0}^{\mathcal{K}}|\langle{r_{j}|\psi(t)\rangle}|^{4}, \hspace{0.2cm}
    \textrm{KIPR}_{\textrm{L}}(t)= \sum_{j=0}^{\mathcal{K}}|\langle{l_{j}|\psi(t)\rangle}|^{4}\,.}
\end{equation}
\textcolor{black}{Now, we proceed to compute these quantities for our model mentioned in (\ref{ham_1}). For our case, $ \textrm{KIPR}_{\textrm{L}}(t)$ and  $\textrm{KIPR}_{\textrm{R}}(t)$ are same. Hence, in all subsequent analyses, we will drop the subscript and simply refer to it as $ \textrm{KIPR}\,.$ We will consider a lattice (unless stated otherwise) consisting of L= 20 unit cells, which consists of a single-particle Hilbert space of dimension 40$\times$40}.
\textcolor{black}{We will also take the values of intra-cell ($w$)and inter-cell hopping ($v$) parameters as $1.5$ and $0.5$ respectively for all numerical calculations throughout the paper.} Starting with the initial state where the particle is localized on the $15^{\textrm{th}}$ site of the lattice, we compute the spread complexity of the evolved state for our model. Below, we summarize our results.
\begin{figure}[htb!]
    \centering    \includegraphics[width=0.45\linewidth]{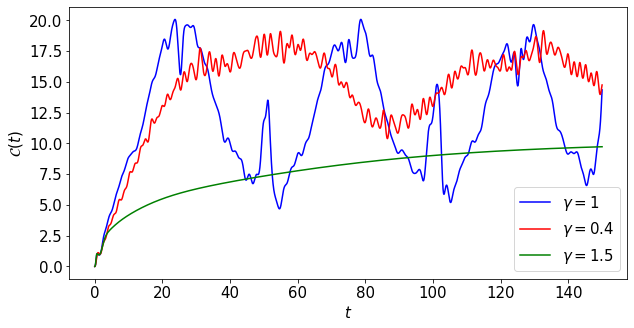}  
    \caption{Time evolution of spread complexity for a non-Hermitian SSH model with $20$ unit cells for different values of $\gamma\,.$}
    \label{Fig-2}
\end{figure}
\begin{itemize}
    \item  It can be seen from Fig.~(\ref{Fig-2}), that in the $\mathrm{PT}$ symmetric phase (when $\gamma$ <1), the spread complexity oscillates with time with a large number of oscillations. At the critical point, i.e., for $\gamma=1\, $, it oscillates, but the number of oscillations decreases. Above the transition point (i.e., $\gamma >1$), the spread complexity saturates after an initial rise. Since the eigen-spectrum is real in the $\mathrm{PT}$ symmetric phase, different phases in the wave function add up to make the spread complexity oscillate. 
    In the $\mathrm{PT}$ broken phase, the eigen-spectrum is imaginary. Hence, the oscillations die down in this region and saturate at a late time. This saturation of the complexity can be attributed to the \textcolor{black}{localization of wave function in the Krylov basis} as reported earlier in \cite{Bhattacharya:2024hto}. The spread complexity seemingly displays two types of different behaviour in these two phases. This is similar to the case reported for the tight-binding model in \cite{Bhattacharya:2024hto}\,. \textcolor{black}{Although the result we have shown here for the non-Hermitian SSH model with open boundary conditions, we have checked that this feature persists (albeit of the fact that the dimension of the Krylov space reduces for periodic boundary conditions) even if we use the periodic boundary condition. We have shown the results for the periodic boundary condition in Appendix~(\ref{app-d}). Furthermore, in Appendix~(\ref{app-c}), we have investigated the behaviour of spread complexity above and below the PT transition point and its saturation value in the PT broken region for different choices of the initial state. We find that its behaviour across the PT transition point remains the same for different choices of the initial state. However, the saturation value above $\gamma=1$ depends on the nature of the initial state as depending on that choice, the value of KIPR changes as shown in Fig.~(\ref{KIPR initial state dependence}) of Appendix~(\ref{app-c}), signaling different amount of localization in the Krylov space. } 

    \begin{figure}[htb!]
    \centering
    \subcaptionbox{$\textrm{Spread Entropy}(\mathcal{S}(t))$}[0.45\linewidth]{\includegraphics[width=\linewidth]{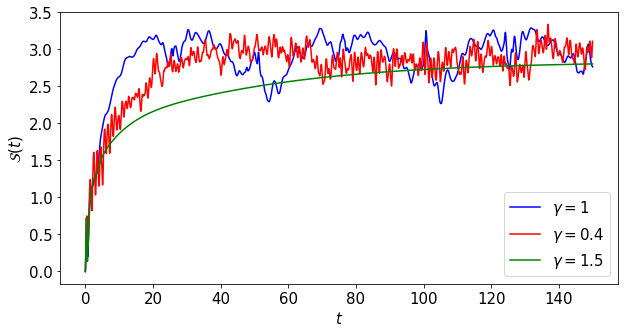}}
    \hfill
    \subcaptionbox{$\textrm{Entropic complexity} (\mathcal{C}_{S}(t))$}[0.45\linewidth]{\includegraphics[width=\linewidth]{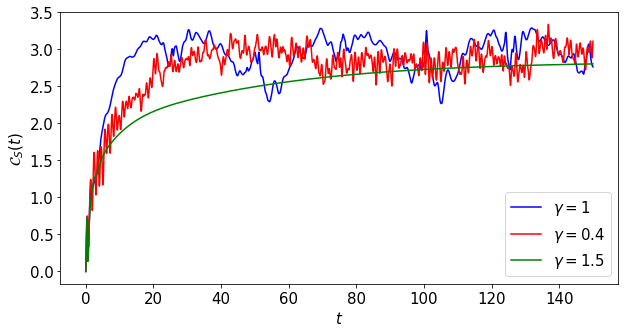}}
    \hfill
    \caption{Time evolution of spread  entropy, and entropic complexity for a non-Hermitian SSH model with $20$ unit cells for different values of $\gamma\,.$}
    \label{Fig-2a}
\end{figure}
    \item In Fig.~(\ref{Fig-2a}), we have displayed the behaviour of the spread entropy, which was defined in (\ref{defination}). The spread entropy also shows a similar kind of behaviour as the spread complexity for the same initial state, and its behaviour can be explained similarly. We have also demonstrated the behaviour of the entropic complexity (defined in  (\ref{definition1})) in Fig.~(\ref{Fig-2a}). Entropic complexity measures the localization strength. The saturation of entropic complexity also reflects the localization of wave function in the $\mathrm{PT}$ broken phase. This supports our earlier observation. 
    \begin{figure}[htb!]
    \centering    \includegraphics[width=0.45\linewidth]{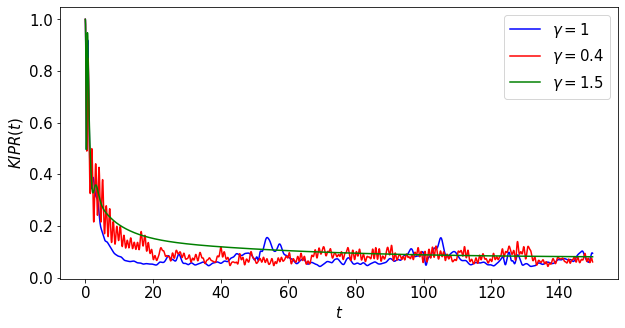}
    \caption{KIPR for a non-Hermitian SSH model with 20 unit cells below and above and below the $\textrm{PT}$ transition point\,.}
    \label{KIPR-PT}
\end{figure}
    \item \textcolor{black}{Furthermore, we have shown the behaviour of the $\textrm{KIPR}$ below and above the $\textrm{PT}$ transition point. It can be seen from the Fig.~(\ref{KIPR-PT}) that at $t=0$, the value of $\textrm{KIPR}$ is one since we start from an initial state, which is fully localized at a single Krylov vector. In the $\textrm{PT}$ symmetric region, when $\gamma < 1$, then $\textrm{KIPR}$ first decays and then oscillates around a small value. The number of oscillations decreases at the $\textrm{PT}$ transition point when $\gamma=1$. In the $\textrm{PT}$ broken region, $\textrm{KIPR}$ first decays and then saturates to a constant value, which is also small, although the oscillations observed in the PT-symmetric region have died down completely. This is perhaps an indication of a localization-delocalization transition in the Krylov space but less sharp than what was observed for the tight-binding model in \cite{Bhattacharya:2024hto}. A more detailed study is required to gain better insight into such localization-delocalization transition, which we leave for the future as our goal in this paper is different, and we are more focused on capturing the entanglement transition.}
\end{itemize}


So far, we have focused on the behaviour of spread complexity and associated quantities across the  $\mathrm{PT}$  transition point. As mentioned earlier, our model also displays a measurement-induced phase transition at $\gamma=2\,$. We will now focus on the spread complexity and entropy across this transition point and analyse their behaviour to see whether we can detect such transition through these quantities. We make the following observations.

\begin{figure}[htb!]
    \centering
    \subcaptionbox{{$\textrm{Spread Complexity}$}}[0.35\linewidth]{\includegraphics[width=1.2\linewidth]{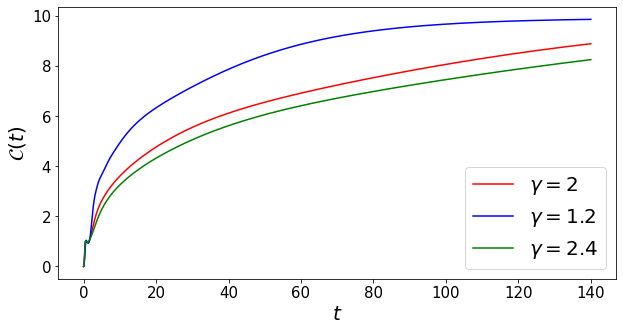}}
    \hspace{0.08\linewidth}
    \subcaptionbox{$\textrm{Spread Entropy}$}[0.40\linewidth]{\includegraphics[width=1.1\linewidth]{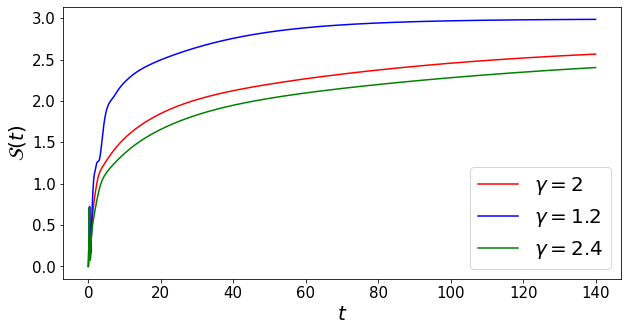}}
    \hspace{0.08\linewidth}
    \subcaptionbox{$\textrm{Entropic complexity}$}[0.40\linewidth]{\includegraphics[width=1.1\linewidth]{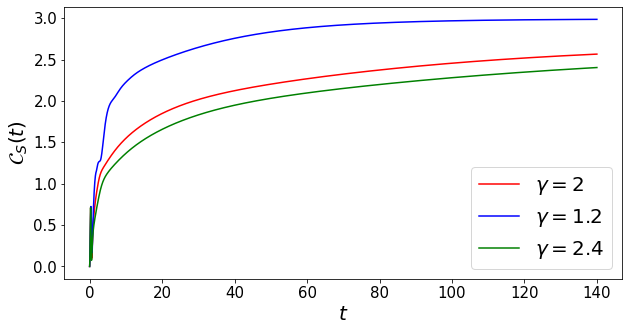}}
    \hspace{0.08\linewidth}
    \caption{Time evolution of spread complexity, spread entropy, and entropic complexity beyond the $\mathrm{PT}$ symmetric region for a non-Hermitian SSH lattice model with 20 unit cells\,.}
    \label{Fig-3}
\end{figure}

\begin{itemize}
\item In Fig.~(\ref{Fig-3}), we have shown that spread complexity saturates above the $\mathrm{PT}$ symmetric region. From Fig.~(\ref{Fig-3}), it can be seen that as the value of the non-hermiticity parameter ($\gamma$) increases, the saturation value of the spread complexity, spread entropy, and entropic complexity decreases and the saturation timescale increases. \textcolor{black}{To better understand the reason behind the lower saturation value of $\mathcal{C}(t)$ for higher values of $\gamma$, one can investigate the Krylov inverse participation ratio ($\textrm{KIPR}$) as defined in \eqref{KIPR eqn}. As shown in Fig.(\ref{Fig-4}), the saturation value of $\textrm{KIPR}$ increases with the increasing value of $\gamma\,.$ A higher value of $\textrm{KIPR}$ indicates stronger localization, and hence, the spread of the state is less. This corroborates the fact the saturation value of the spread complexity decreases with the increasing value of $\gamma$ (in the PT broken region), irrespective of the entanglement transition point. This is similar in line with what has been observed for  \cite{Bhattacharya:2024hto} for the tight-binding model.}

\begin{figure}[htb!]
    \centering
    \includegraphics[width=0.50\linewidth]{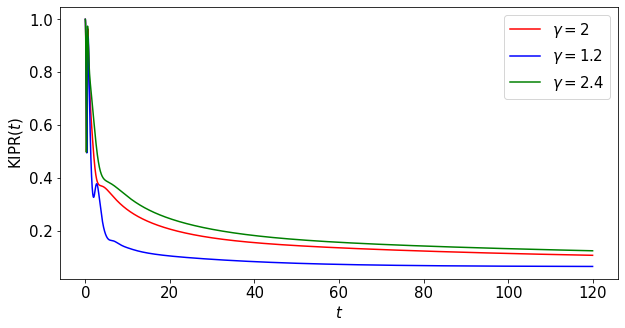}
    \caption{KIPR for a non-Hermitian SSH model with 20 unit cells beyond the PT symmetric region for different values of $\gamma$ in the PT broken region\,.}
    \label{Fig-4}
\end{figure}
\item To clarify things, we plot the saturation timescale of spread complexity and entropy as a function of $\gamma$ in the PT broken phase, especially around the entanglement transition point in Fig.~(\ref{Fig-5}). We find that the saturation time scale increases with the increase of the parameter (measurement rate) $\gamma$, but it is insensitive to entanglement transition.
 
 \begin{figure}[htb!]
    \centering
    \subcaptionbox{Saturation time scale of spread complexity}[0.45\linewidth]{\includegraphics[width=\linewidth]{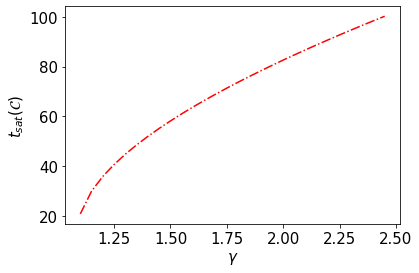}}
    \hfill
    \subcaptionbox{Saturation time scale of entropic complexity}[0.45\linewidth]{\includegraphics[width=\linewidth]{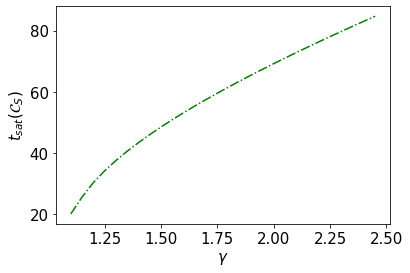}}
    \hfill
    \caption{Saturation timescale of the spread complexity and entropic complexity as a function of measurement rate $\gamma\,$ for anon-Hermitian SSH Lattice Model with 20 unit cells. }
    \label{Fig-5}
\end{figure}

\item Keeping in mind that the measurement-induced transition can be detected by looking at the scaling of entanglement entropy with respect to (sub) system size \cite{PhysRevB.103.224210,10.21468/SciPostPhysCore.6.3.051},  we investigate the scaling of the spread complexity with respect to the system size ($L$) around this transition point (i.e around $\gamma=2$)\,.
\begin{figure}[htb!]
    \centering
    \includegraphics[width=0.5\linewidth]{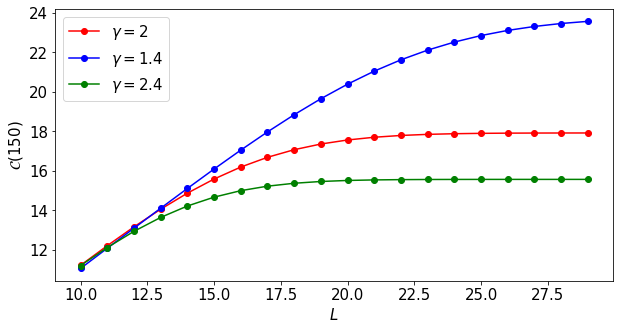}
    \caption{Late-time spread complexity for different system size  $L\,.$}
    \label{Fig-6}
\end{figure}
As can be seen from Fig.~(\ref{Fig-6}), the spread complexity initially increases following a saturation for a larger system size ($L$). The initial increase in spread complexity decreases with $\gamma$, and the saturation value decreases with measurement rate $\gamma$.  
Also, from Fig.~(\ref{Fig-6}), it seems that there is a change in the scaling behaviour of the spread complexity from linear to sub-linear as we increase the value of the $\gamma\,.$ To make this observation concrete, we have checked the scaling of spread complexity: $\mathcal{C} \propto L^{\alpha}$ and found that $\alpha$ smoothly decreases across the $\gamma=2$ without showing any jump.
The scaling factor($\alpha(\gamma)$) is unable to detect the entanglement transition point ($\gamma=2$). The reason behind this failure to detect entanglement transition could be the absence of information of any subsystem. Unlike the entanglement entropy for a subsystem, we are looking at the complexity of the state for the entire system. Keeping this in mind; in the next section, we will extend our investigation of the spread complexity for a subsystem and try to see whether it can capture this \textcolor{black}{measurement-induced} transition.
\end{itemize}
\textit{Spread Complexity for a subsystem  in Krylov Space:} \\\\
We now turn our attention to the complexity of a subsystem in the Krylov space. Starting with an initially localized state in real space, we applied the bi-Lanczos algorithm for our Hamiltonian model (\ref{ham_1}). We expand the initial state $|\psi(0)\rangle$ in terms of Krylov basis vectors and evolve the state using the procedure explained in Appendix~(\ref{app-B}). After that, we construct the density matrix $\rho(t)$  as mentioned in (\ref{krylov space rho}) and expand in terms of Krylov basis vectors. From this, we calculate the reduced density matrix $\rho_{\textrm{A}}=\Tr_{\textrm{B}}\rho$ by taking the reduced trace over a subspace in Krylov space.
Now, we like to compute the spread complexity for this $\rho_{\textrm{A}}\,.$ Note that $\rho_{\textrm{A}}$ corresponds to a mixed state. Then, taking a cue from the measures for entanglement for a mixed state such as  ``Entanglement of Purification''($\mathrm{EoP}$) and a similar measure for circuit complexity such as ``Complexity of Purification'' ($\mathrm{CoP}$) \cite{Nguyen:2017yqw, PhysRevLett.122.201601, PhysRevResearch.3.013248,Bhattacharyya:2020iic,Bhattacharyya:2021fii,Bhattacharya:2022wlp,Bhattacharyya:2022rhm}  we use the Choi–Jamiołkowski isomorphism to purify this $\hat{\rho}_A$ to get a pure state in the doubled Hilbert space. An operator $\hat{O}$  can  be expanded in terms of a orthonormal  basis $|i\rangle$ in the following way: $$\hat{O}=\sum_{i\,,j} O_{ij}|i\rangle\langle j|\,.$$ Then following \cite{PhysRevA.87.022310} 
one can associate a state with this operator via channel-state duality/Choi–Jamiołkowski isomorphism,  \begin{equation} \label{channel-state}
\hat{O}= \sum_{i\,,j} O_{ij}|i\rangle\langle j| \leftrightarrow |\hat{O}\rangle =\frac{1}{\sqrt{\Tr(O^{\dagger}O)} }\sum_{i,j}\,O_{ij}|i\rangle \otimes |j\rangle\,.
\end{equation}
This idea can be used to purify the reduced density matrix $\hat{\rho}_A$. We have expanded  $\hat{\rho}_A$ in terms of its eigenbasis (which forms an orthonormal basis). $$\hat{\rho}_{A}=\frac{1}{\sqrt{\sum_j\lambda^2_j}}\sum_{i}\lambda_i|\lambda_{i}\rangle\langle \lambda_i|$$ where $\lambda_is$ denote the eigenvalues and $|\lambda_i\rangle$ are the eigenvectors. After that, using the channel-state duality we get, \begin{equation} |\hat{\rho}_A\rangle=\frac{1}{\sqrt{\sum_j\lambda^2_j}}\sum_{i}\lambda_i|\lambda_{i}\rangle\otimes |\lambda_i\rangle\,. \label{pur1}\end{equation}
Then one can use the same procedure outlined at the beginning of this section to compute the spread complexity of purification ($\textrm{kCoP}$) using the bi-Lanczos coefficients generated earlier \footnote{In recent times, a version of it has been explored for certain two-qubit states as well as for certain coherent states in \cite{Das:2024zuu}\,.}\,.\par 
\begin{itemize}

\item 
First, we  analyze the scaling of late-time 
 $\textrm{kCoP}$ with sub-system size. \textcolor{black}{For this purpose, we have considered $L=200$ unit cells and then find the late-time $\textrm{kCoP}$ for different sub-system sizes ($l=4,8,10,20,40,50,100$)}. It takes the following form $$\textrm{kCoP}\propto \ell^{\alpha(\gamma)}\,.$$
\begin{figure}[htb!]
    \centering
    \includegraphics[width=0.50\linewidth]{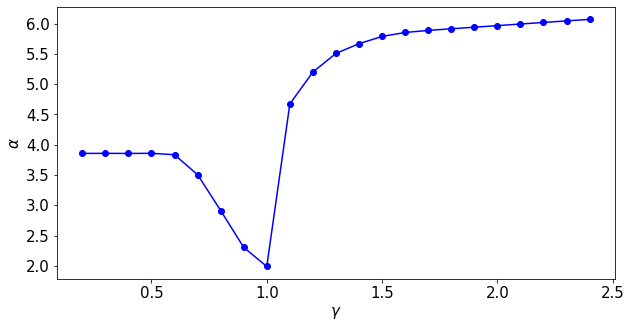}
    \caption{Behaviour of the scale factor associated with the power law scaling of late time \textrm{kCoP} with $\gamma\,.$}
    \label{Fig-7}
\end{figure}\\
Fig.~(\ref{Fig-7}) shows the variation of $\alpha$ with respect to the measurement rate $\gamma\,.$ It shows a dip at $\gamma=1$, which is the $\mathrm{PT}$ transition point, then sharply increases before reaching a plateau. So we can see that, using the late time scaling of the $\textrm{kCoP}$, we can detect the $\mathrm{PT}$ transition point quite effectively, although it fails to detect the measurement-induce transition point ($\gamma=2$). 

\item Recently, it has been shown in \cite{PhysRevB.109.224304} that the time-averaged Krylov complexity can act as an order parameter for phase transitions for a certain model (e.g., the Lipkin-Meshkov-Glick model). Motivated by this idea, we now investigate the time averaged  $\textrm{kCoP}$ to check whether it can be an order parameter for our case. The time average  $\textrm{kCoP}$ is defined as,
\begin{equation}\label{avg C_{K} equ}
    \bar{C}_{K}= \frac{1}{T - \textrm{tref}}\int_{\textrm{tref}}^{T} C_{K}(t)\,dt\,.
\end{equation}
\textcolor{black}{In Fig.~(\ref{Fig-8})}, we plot  $ \bar{C}_{K}$  as a function of measurement rate $\gamma\,.$
\begin{figure}[htpb!]
    \centering
    \includegraphics[width=0.60\linewidth]{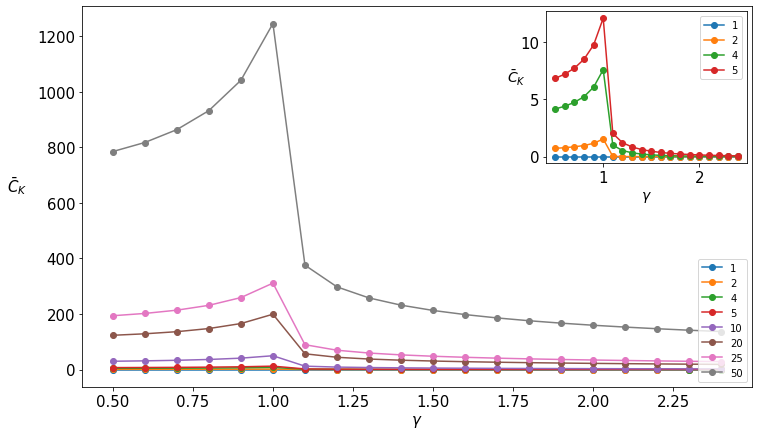}
    \caption{$\bar{C}_{K}$ vs $\gamma$ for different sub-system size for a non-Hermitian \textrm{SSH} lattice model with $L=100$ unit cell\,. Inset: $\bar{C}_{K}$ vs  $\gamma$ for small sub-systems.}
    \label{Fig-8}
\end{figure}
We can see a sharp transition around the $\textrm{PT}$ transition point, i.e., $\gamma=1$, and it starts to become independent of $\gamma$ for higher values of $\gamma$ irrespective of the measurement-induced transition point. \textcolor{black}{This phenomenon is prominent even if the sub-system size is small}. 
We also explore the scaling of $\bar{C}_{K}\propto$ $\ell^{\alpha}$ and plot $\alpha$ as a function of $\gamma$ in Fig.~(\ref{Fig-9}). 
It can be seen that it changes sharply near $\gamma=1$, which is the $\textrm{PT}$ transition point. 
\begin{figure}[htpb!]
    \centering
    \includegraphics[width=0.6\linewidth]{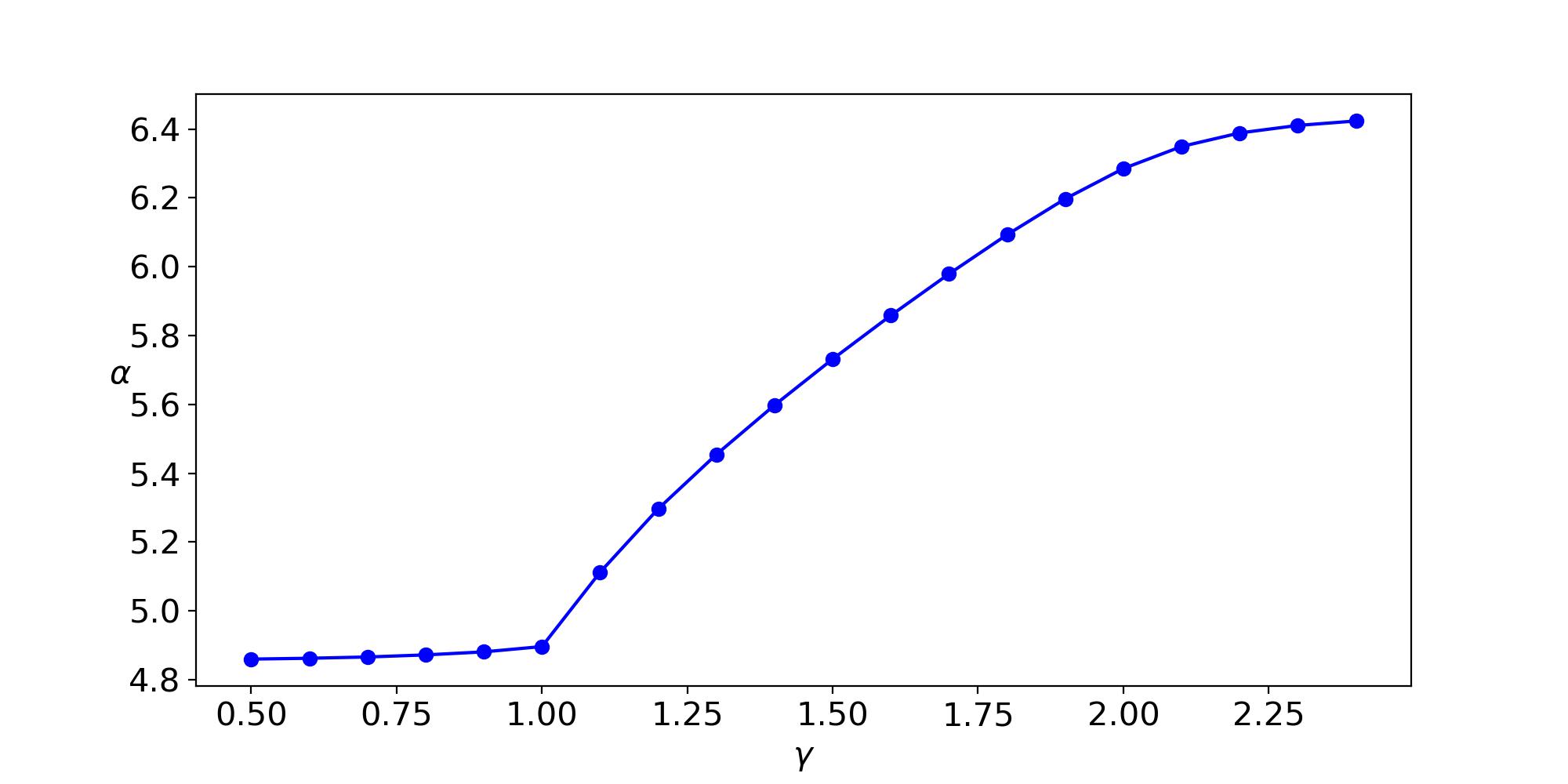}
    \caption{Scaling of $\bar{C}_{K}$ as a function of $\gamma\,$ for a non-Hermitian SSH lattice Model with 100 unit cells\,.}
    \label{Fig-9}
\end{figure}
\end{itemize}
So we conclude that the scaling of the late time value of the $\mathrm{kCoP}$ (with respect to subsystem size) can serve as a good indicator (in fact, better than the spread complexity evaluated earlier for the whole system) for the $\mathrm{PT}$ transition, as its behaviour changes sharply across the transition point, but still $\mathrm{kCoP}$ cannot probe the measurement-induced transition point. 
It is still insensitive to the correlation between the subsystem and its complement in real space, unlike the entanglement entropy. 
\par
\textcolor{black}{Before ending this section, to get an additional insight into the behaviour of the scaling of late-time kCoP as shown in Fig.~(\ref{Fig-7}), we investigate the   $\textrm{KIPR}$ for the purified state (in \eqref{pur1}) across the $\textrm{PT}$ transition\footnote{\textcolor{black}{Although we have shown the result for one particular sub-system ($l=20$), we checked that the qualitative behaviour remains the same for different sub-system sizes.}}}.
\begin{figure}[H]
    \centering    \includegraphics[width=0.60\linewidth]{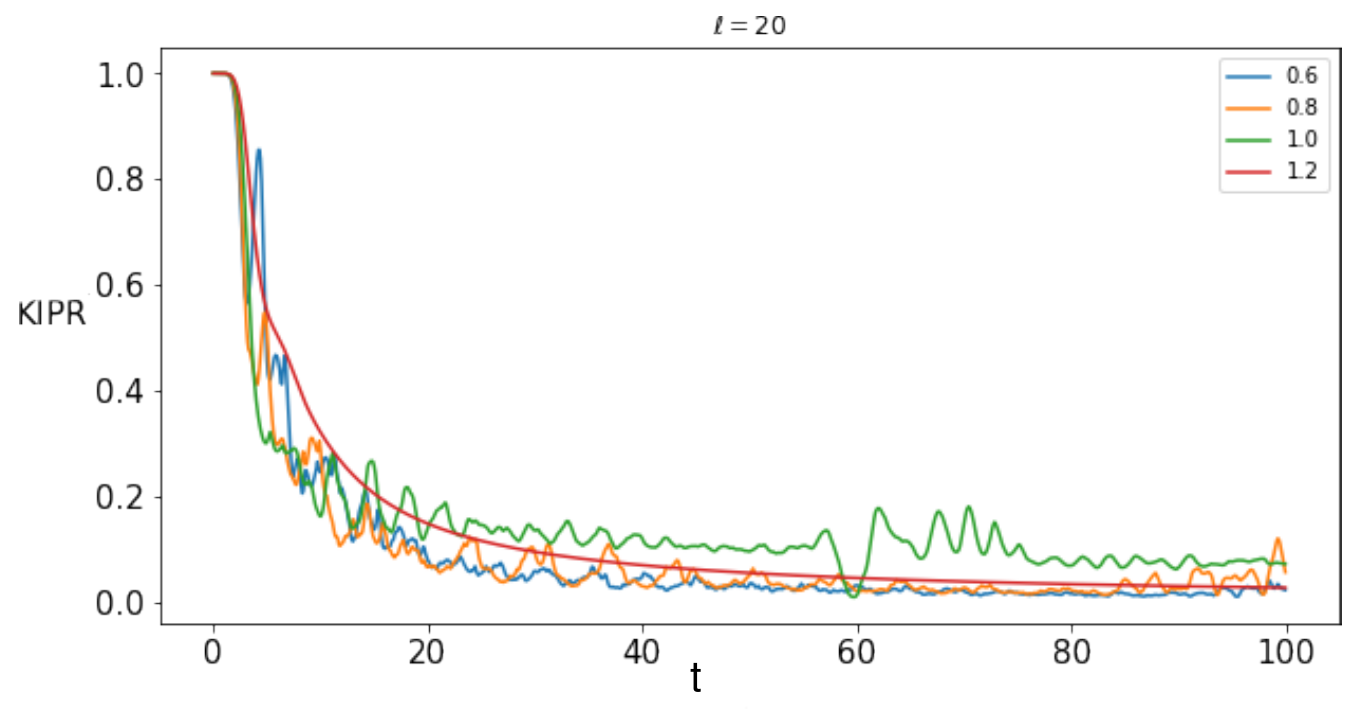}
    \caption{$\textrm{KIPR}$ for the purified state. System size is  $L= 100$ unit cells and the sub-system size is $\ell=20$  for 4 different $\gamma$ values.} 
    \label{KIPRP}
\end{figure}
\textcolor{black}{From Fig.~(\ref{KIPRP}), it is evident that when $\gamma \leq 1$  the $\textrm{KIPR}$ is oscillating (in fact it increases at $\gamma=1$) and just above $\gamma=1$ value, it saturates. This indicates an abrupt change from delocalization to localization in the Krylov space. The sudden dip in the scaling exponent of $\textrm{kCoP}$ as shown in Fig.~(\ref{Fig-7}) can be attributed to this fact since it is coming from the late-time value of the spread complexity for this purified state, which depends on  localization-delocalization}. \par
\begin{figure}[htb!]
    \centering
    \includegraphics[width=0.50\linewidth]{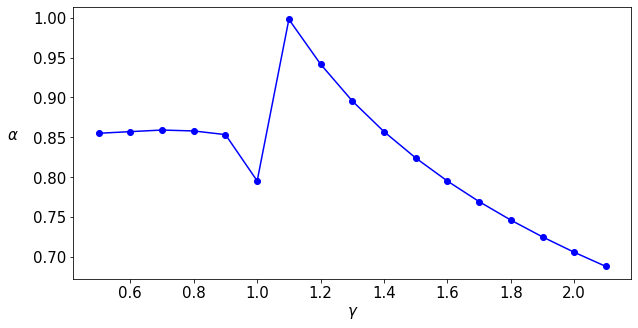}
    \caption{Scale factor $\alpha$ associated with the power law scaling of $\overline{\textrm{KIPR}}$ for the purified state for $l=20$ (sub-system size)\,.} 
    \label{alphavsgamma_KIPRP_}
\end{figure}

 \textcolor{black}{Further, we have investigated the scaling of the time-averaged ${\textrm{KIPR}}$ with respect to the system-size and found that it also follows a power law behaviour which is of the form: $\overline{\textrm{KIPR}} \propto \ell^{-\alpha(\gamma)}\,.$ From the Fig.~(\ref{alphavsgamma_KIPRP_}), we can easily see that it shows a sudden change precisely across $\gamma=1$ (like what happens for the kCoP as shown in Fig.~(\ref{Fig-7})), further corroborating our above-mentioned reasoning}. 
\newpage
\par
In the next section, we focus on another quantity, namely \textit{Quantum Fisher Information}, which, when evaluated in the Krylov space,  possesses a structural similarity to the Krylov complexity \cite{Shi:2024bpu, PhysRevLett.133.110201}. Motivated by this fact, we will explore this quantity to see whether it can detect the entanglement transition.
\section{Quantum Fisher information in Krylov space}\label{section-6}
Quantum Fisher information (\textrm{QFI}) is a central object in quantum measurement and quantum metrology. The classical Fisher information describes how rapidly a probability distribution changes with respect to some parameters. Similarly, \textrm{QFI} describes how fast a quantum state changes with respect to some parameter. \textrm{QFI} has been used to detect quantum phase transition \cite{Wang_2014, PhysRevB.100.184417} as well as the scaling of time-averaged \textrm{QFI} has  been used as a diagnostics of  quantum chaos \cite{Shi:2024bpu}. Recently,  it has been shown that the time-averaged \textrm{QFI} can be used to detect the measurement-induced phase transition for models (collective spin systems) \cite{Poggi:2023mll}. Furthermore, from the analysis of \cite{Shi:2024bpu, PhysRevLett.133.110201},  that  $\textrm{QFI}$ possess a structural similarity with the Krylov complexity when decomposed in terms of Krylov basis. We will elaborate more on this shortly. Motivated by these, we investigate the time-averaged \textrm{QFI} in Krylov space for our model to possibly detect the measurement-induced entanglement transition in the $\textrm{PT}$ broken region.\par 
The $\textrm{QFI}$ for a pure state $|\psi(t)\rangle$ can be defined for an operator $\hat{O}$ in the following way \cite{Pezze:2016nxl, 2014JPhA...47P4006T, Chu:2023mzh,Shi:2024bpu, PhysRevLett.133.110201},
\begin{equation}\label{QFI Pure state}
    F_{Q}[\ket{\psi(t)},\hat{O}]= 4 (\Delta\hat{O}(t))^{2})= 4 (\bra{\psi(t)}\hat{O}^{2}\ket{\psi(t)}-\bra{\psi(t)}\hat{O}\ket{\psi(t)}^{2}).
\end{equation}
where $\ket{\psi(t)}$ is given by equation ($\ref{Evolpe eqn}$).\\
The time-averaged $\textrm{QFI}$ can then be defined same as in (\ref{avg C_{K} equ}),
\begin{equation}\label{averaged QFI}
    \textcolor{black}{\bar{F}_{Q}}=\frac{1}{T-t_{ref}}\int_{t_{ref}}^{T} F_{Q}[\ket{\psi(t)},\hat{O}]\,dt\,.
\end{equation}
For our case, we will focus on studying this quantity in Krylov space. We do it in two ways: firstly, by expanding the state in terms of the Krylov basis and, secondly, by expanding the operator in the Krylov space. 
\subsection*{Krylov expansion of state}
The time-evolved state $|\psi(t)\rangle$ can be expanded in Krylov basis as, 
\begin{equation}
    \ket{\psi(t)} = \sum_{j=1}^{\mathcal{K}}\Psi_{j}^{r}(t)\ket{r_{j}}= \sum_{j=1}^{\mathcal{K}}\Psi_{j}^{l}(t)\ket{l_{j}} \,.
\end{equation}
The expectation values of the operators can be expanded in terms of  the Krylov basis as
\begin{equation}
    \langle{O^{k}(t)}\rangle = \sum_{j,j'=1}^{\mathcal{K}}{\Psi_j^{l*}(t)\Psi_{j'}^r(t)\bra{l_{j}}O^{k}\ket{r_{j}'}}\,.
\end{equation}
For our non-Hermitian \textrm{SSH} model the measurement operators are $\hat{n}_{A}$ for sub-lattice $A$ and $I-\hat{n}_{B}$ for sub-lattice $B$. We have computed time-averaged \textrm{QFI} using (\ref{averaged QFI}) for the measurement operators $\hat{n}_{A}$ and $I-\hat{n}_{B}$ for different system sizes and for different measurement rate $\gamma\,.$
\begin{figure}[H]
    \centering
    \includegraphics[width=0.8\linewidth]{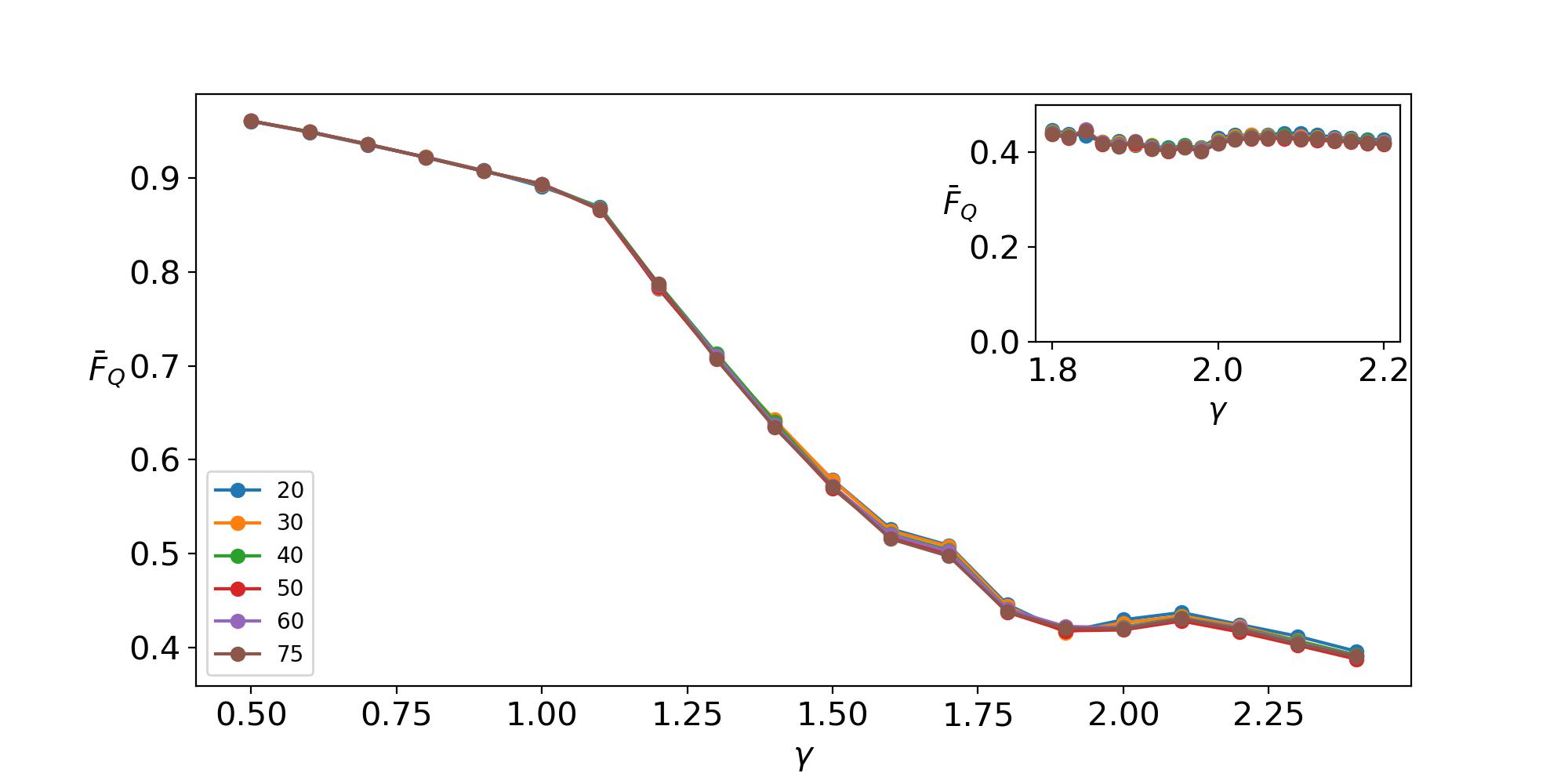}
    \caption{Time-averaged QFI as a function of $\gamma $ for different system size ($L$) as indicated by the legends in the inset. In the inset, we have shown the behaviour of QFI around the measurement-induced transition point, i.e., $\gamma=2\,.$}
    \label{Fig-10}
\end{figure}
Now we make the following observations:
\begin{itemize}
    \item 
From Fig.~(\ref{Fig-10}), it is evident that the nature of the $\bar{F}_{Q}$ is \textit{insensitive to the system size}\,. 
\item Secondly, it can be seen from Fig.~(\ref{Fig-10}), that $\bar{F}_{Q}$ decreases with increasing value of $\gamma$. There is a change of its slope at $\gamma=1$ ($\textrm{PT}$ transition point), and it almost becomes independent of $\gamma$ around $\gamma=2$. We have shown the behaviour of $\bar{F}_{Q}$ near $\gamma=2$ in the inset of Fig.~(\ref{Fig-10}). So we can conclude that the \textit{time-averaged $\textrm{QFI}$ can \textcolor{black}{probe the measurement-induced phase} in our model besides probing the  $\textrm{PT}$ transition \footnote{In recent times, QFI has been used to probe certain monitored Clifford circuits and collective spin systems. Interested readers are referred to \cite{Poggi:2023mll,Lira-Solanilla:2024alp}.}.} 
\end{itemize}
\subsection*{Krylov expansion of the operator}
Next, we will use the Heisenberg picture to write the $\textrm{QFI}$ for a pure state in the following way,
\begin{equation}
    F_{Q}(\rho(t),\hat{O})= 4\Bigg(\frac{1}{N^{2}}\bra{\psi(0)}\hat{O}^{2}(t)\ket{\psi(0)}- \frac{1}{N^{4}}\bra{\psi(0)}\hat{O}(t)\ket{\psi(0)}^{2}\Bigg)
\end{equation}
where $\hat{O}(t)$ (for a generic non-Hermitian $H$) can be expressed as,
\begin{equation}\label{operator equation non hermitian}
    \hat{O}(t)= e^{iH^{\dag}t}\hat{O}(0)e^{-iHt}\,,\quad N= ||e^{-iHt}\ket{\psi(0)}||\,.
\end{equation}
As mentioned previously, $\hat{O}(0)$ for our case will be the measurement operator $\hat{n}_{A}$ and $I-\hat{n}_{B}$ and the Hamiltonian is given by (\ref{ham_1}). Since our Hamiltonian is non-Hermitian, we will use the bi-Lanczos method 
(extending it for the operators) to find the left and right Lanczos coefficients and the left and right Krylov basis. The details are given in Appendix~(\ref{app-A}). 
Then we can expand the operator $\hat{O}(t)$ as, 
\begin{align}\label{Krylov expansion}
\begin{split}
    \hat{O}(t)=\sum_{m} i^{m} \phi_{m}^{R}(t) \hat{O}_{r_{m}}\,,\\
    =\sum_{n} i^{n} \phi_{n}^{L}(t)\hat{O}_{l_n}
\end{split}
\end{align}
where $\phi_{m}^{R}(t)$ and $\phi_{m}^{L}(t)$ are the operator wave function and $\hat{O}_{r_{m}},\hat{O}_{l_{n}}$ are the right and left Krylov basis as defined in (\ref{Appeq1}).
For our case using the expansion (\ref{Krylov expansion}) we can re-write the expression for $\textrm{QFI}$ as \footnote{For the non-Hermitian case, we have also replaced $\langle{\hat{O}^{2}}\rangle\rightarrow\frac{1}{2}(\hat{O}^{\dag}\hat{O} + \hat{O} \hat{O}^{\dag})$ and $\langle{\hat{O}}\rangle^2 \rightarrow \langle{\hat{O}^{\dagger}}\rangle \langle{\hat{O}}\rangle\,.$},
\begin{align}\label{QFIop}
\begin{split}
  F_{Q}(\rho(t),\hat{O})=4\Bigg (\sum_{m,n}  & \frac{i^{m+n}}{N^{2}}\Big\{(-1)^{m} \phi_{m}^{R *}(t)\phi_{n}^{L}(t)\langle{O_{r_{m}}^{\dag} O_{l_{n}}\rangle} + (-1)^{n} \phi_{m}^{R}(t)\phi_{n}^{L *}(t)\langle{O_{r_{m}} O_{l_{n}}^{\dag}\rangle}\Big\}
    - \\& \frac{i^{m+n}}{N^{4}}\Big\{\phi_{m}^{*R}(t) \phi_{n}^{L}(t)\langle{O^{\dag}_{r_{m}}}\rangle\langle{O_{l_{n}}}\rangle\Big\}\Bigg)\,.
\end{split}
\end{align}
We have checked that, for our case, $\phi_{m}^{R *}(t)\phi_{n}^{L}(t)= \phi_{m}^{R}(t)\phi_{n}^{L *}(t)\,.$
Finally the expression for $\textrm{QFI}$ can be re-written as,
\begin{align}\label{QFIop1}
  F_{Q}(\rho(t),\hat{O})=4\Bigg (\sum_{m,n} \phi_{m}^{R *}(t)\phi_{n}^{L}(t) \frac{i^{m+n}}{N^{2}}\underbrace{\Big\{(-1)^{m} \langle{O_{r_{m}}^{\dag} O_{l_{n}}\rangle} + (-1)^{n} \langle{O_{r_{m}} O_{l_{n}}^{\dag}\rangle}-\frac{1}{N^{2}}\langle{O^{\dag}_{r_{m}}}\rangle\langle{O_{l_{n}}}\rangle\Big\}\Bigg)}_{f_{(m,n)}}\,.
\end{align}

It is evident form (\ref{QFIop1}), $\textrm{QFI}$ possesses a structural similarity with complexity functional. Note that, for a unitary evolution, the normalization factor, $N=1$ and
\begin{align}
\begin{split}
  \phi_{m}^{R*}(t)=\phi_{m}^{R}(t)=\phi_{m}(t)\,,\quad 
  \phi_{n}^{L *}(t)=\phi_{n}^{L}(t)=\phi_{n}(t)\,,\quad
  O_{r_{m}}^{\dag}=O_{m},O_{l_{n}}=O_{n}\,.
\end{split}
\end{align}
Then the expression for the $\textrm{QFI}$ in Krylov space as mentioned (\ref{QFIop}) reduces to,
\begin{equation} \label{QFI}
    F_{Q}(\rho(t),\hat{O})= 4\sum_{m,n} i^{m+n} \phi_{m}(t)\phi_{n}(t)
    \underbrace{\Big(((-1)^{m}+(-1)^{n})\langle O_{m}O_{n}\rangle - \langle{O_{m}}\rangle \langle{O_{n}}\rangle\Big)}_{g_{(m,n)}}\,.
\end{equation}
In \cite{PhysRevLett.133.110201}, authors came up with a variant of the QFI using (\ref{QFI}) in the correlation landscape to qualitatively capture dynamical behaviour, which is proportional to the Krylov complexity. In general, one can think of (\ref{QFI}) as a generalization of the complexity functional encapsulating the information of the underlying correlation. For non-unitary evolution, this further generalizes to (\ref{QFIop1}) as mentioned earlier. \par
Now motivated by these facts, following our previous analysis regarding Krylov complexity, we now calculate the time-averaged $\textrm{QFI}$ as defined in (\ref{averaged QFI}) in Krylov space for different system sizes for various values of measurement rate $\gamma$. We have observed that, under time-averaging, the  contribution from the off-diagonal part of (\ref{QFIop1}) is negligible. Hence we construct the following by keeping only the diagonal part (i.e, $n=m$ terms) from (\ref{QFIop1}), 
\begin{align}
  F_{Q}^{\textrm{diag}}(\rho(t),\hat{O})=\sum_{n} \phi_{n}^{R *}(t)\phi_{n}^{L}(t)\underbrace{\frac{4}{N^{2}}\Big\{\langle{O_{r_{n}}^{\dag} O_{l_{n}}\rangle} +  \langle{O_{r_{n}} O_{l_{n}}^{\dag}\rangle}-\frac{(-1)^n}{N^{2}}\langle{O^{\dag}_{r_{n}}}\rangle\langle{O_{l_{n}}}\rangle\Big\}\Bigg)}_{f_{n}}\,.  \label{QFIop2}
\end{align}
Now if we compare (\ref{QFIop2}) with the complexity functional $ C_{K}=\sum_{n} n\, \phi_{n}^{R *}(t)\phi_{n}^{L}(t)\,,$ we can easily see that, (\ref{QFIop2}) has a \textcolor{black}{structural similarity with it}. 
It incorporates the information of the correlation through $f_n$ and hence possess a much richer structure. \par
Now, we will explore both the time-averaged QFI coming from the (\ref{QFIop1}) as well as the generalized complexity measure coming from (\ref{QFIop2}). Along with that we have also plotted the diagonal part of the $\textrm{QFI}$ separately.

 \begin{figure}[htb!]
    \centering
    \subcaptionbox{$\bar{F}_{Q}$ vs $\gamma$}[0.46\linewidth]{\includegraphics[width=0.90\linewidth] 
    {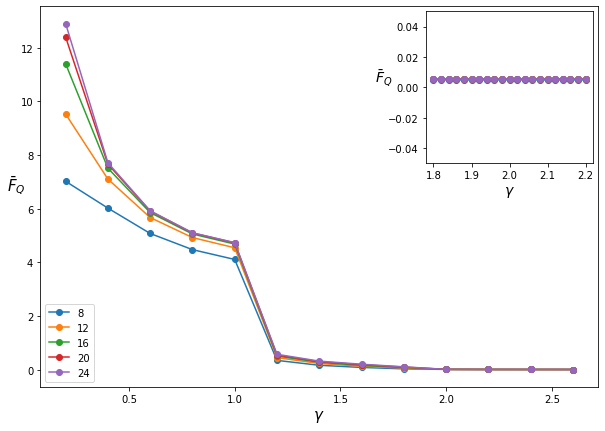}} \label{Fig-12a}
    \hfill
    \subcaptionbox{$\bar{F}^{\textrm{diag}}_{Q}$ vs $\gamma$}[0.46\linewidth]{\includegraphics[width=1.07\linewidth]{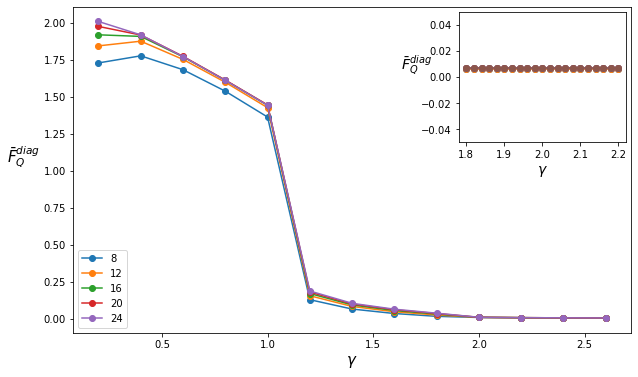}}    \label{Fig-12b}
    \hfill
   \caption{Time-averaged QFI (left) and its diagonal part (right) as a function of $\gamma $ for different system sizes (indicated via legends in the inset). In the inset, we have shown the behaviour of QFI around the measurement-induced transition point, i.e. $\gamma=2\,.$} 
    \label{Fig-12}
\end{figure}


Now we summarize our observations:
\begin{itemize}
    \item It is evident from the Fig.~(\ref{Fig-12}) (left panel) that the behaviour time-averaged QFI  ($\bar{F}_{Q}$) is independent of the system size (except for small values of $\gamma$). Initially, it starts from a higher value, and its slope changes sharply across the PT transition point, i.e., $\gamma=1\,.$
    \item Furthermore, it saturates and becomes vanishingly small across the measurement-induced transition point, i.e. $\gamma=2\,.$ This has been highlighted in the inset of the Fig.~(\ref{Fig-12}) (left panel)\,. \textcolor{black}{Note that we have used the open-boundary condition. These features remain the same if we work with periodic boundary conditions, as shown in Appendix~(\ref{app-e}).}
    \item Also, from the right panel of Fig.~(\ref{Fig-12}), it is evident that the (time-averaged) diagonal part of $\textrm{QFI}$ is able to capture the information about the $\textrm{PT}$ transition as \textcolor{black}{well as probe the $\textit{measurement-induced phase}$} like the full time-averaged $\textrm{QFI}$. It changes slope \textcolor{black}{around} $\gamma=1$, and then it saturates around $\gamma=2$, which we have shown in the inset of Fig.~(\ref{Fig-12}) (right panel) \footnote{\textcolor{black}{At this point, we are unable to pinpoint the exact transition through the behavior of time-averaged QFI (or its diagonal part) as the slop of its starts to saturate a bit before $\gamma=2$ and becomes independent of the sub-system size. Nevertheless it stays very low inside the measurement-induced `area law' phase compared to the `volume law' regime. We leave it for future studies to pinpoint this transition point.}}. \textcolor{black}{Note that, in \cite{10.21468/SciPostPhys.14.5.138}, it was observed that for this non-Hermitian SSH model, the entanglement entropy (in real space) per unit length of the sub-system displays a discontinuous change across the PT transition point (i.e. around $\gamma=1$) inside the `volume law' regime. Then, it further changes discontinuously at $\gamma=2$ to go to an `area law' phase. Now, given the fact that one can think of (time) averaged QFI as a witness of multipartite entanglement, it is not unlikely that it will be sensitive to both of these changes of the entanglement. Hence, this gives a plausible physical reason behind its sharp change at the $\gamma=1$ (as well as around $\gamma=2$)\,. Although much more investigation is needed to gain a deeper insight into this behaviour of time-averaged QFI by extending the similar line of study presented in \cite{Lira-Solanilla:2024alp} for our case first in real space and then in Krylov space to find more physical insight into the observations made in this paper.}
\end{itemize}
Finally, we can conclude that the QFI (time-averaged), which is of similar structure to that of the Krylov complexity when expanded in terms of Krylov basis but captures the information of the underlying correlation (which is missing from the Krylov complexity functional), can be a promising \textit{probe of the measurement-induced phase along with being a probe for $\textrm{PT}$ transition}. \textit{Not only that, the diagonal part it, \textcolor{black}{which is structurally similar to the Krylov complexity functional} but contains information about the correlation landscape as discussed earlier, is enough to capture the information about these two types of phases.}\par
\begin{figure}[H]
    \centering
    \captionsetup{font=small}
    \subcaptionbox{$\gamma=0.4$}[0.35\linewidth]{\includegraphics[width=1.1\linewidth]{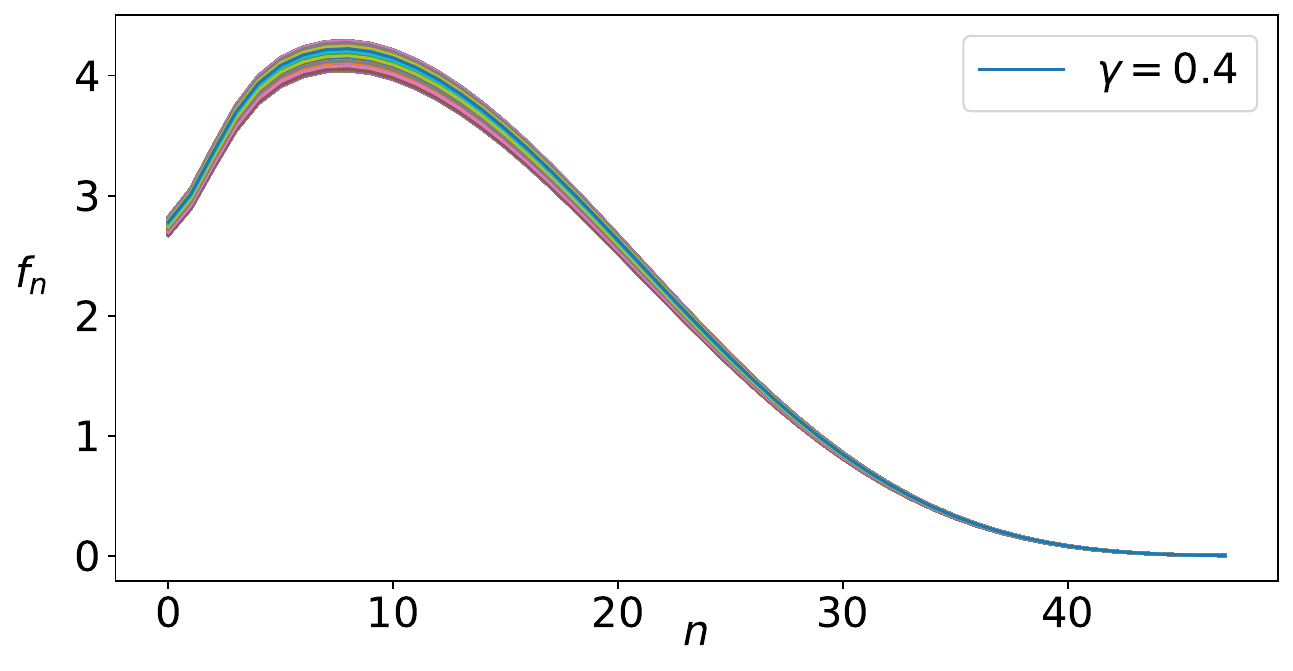}}  
    \hspace{0.04\linewidth}
    \subcaptionbox{$\gamma=1.0$}[0.35\linewidth]{\includegraphics[width=1.1\linewidth]{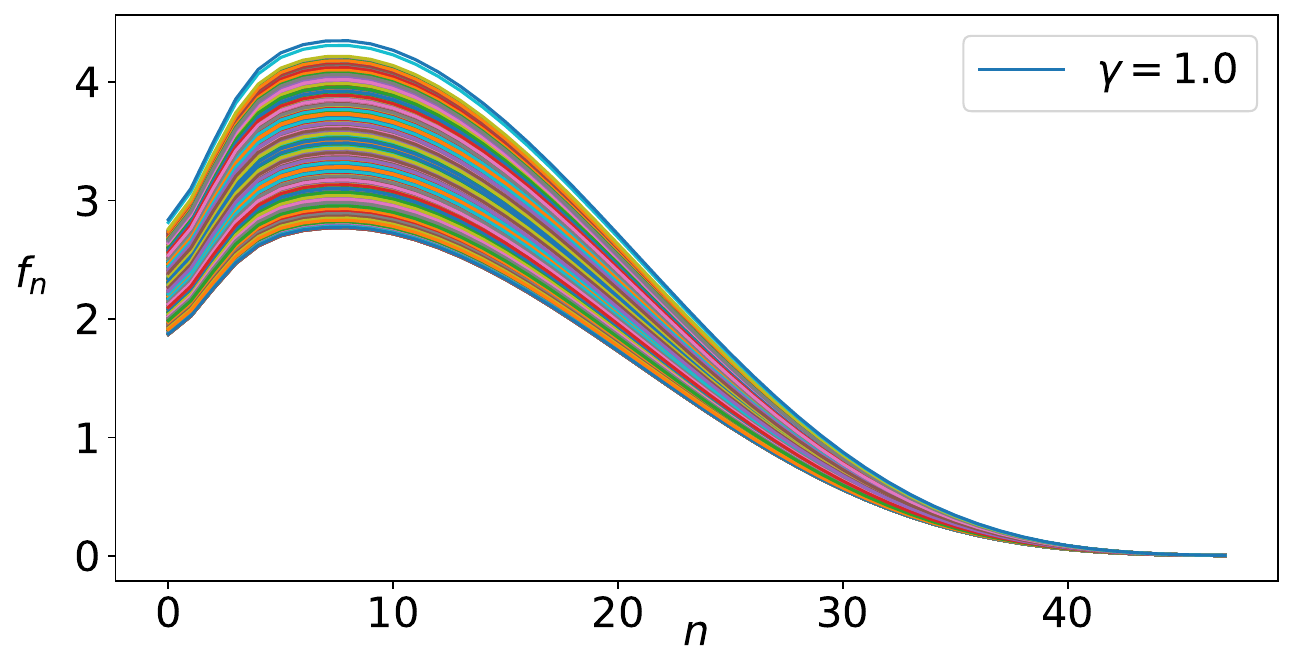}}
    \hspace{0.04\linewidth}
    \subcaptionbox{$\gamma=2.0$}[0.35\linewidth]{\includegraphics[width=1.1\linewidth]{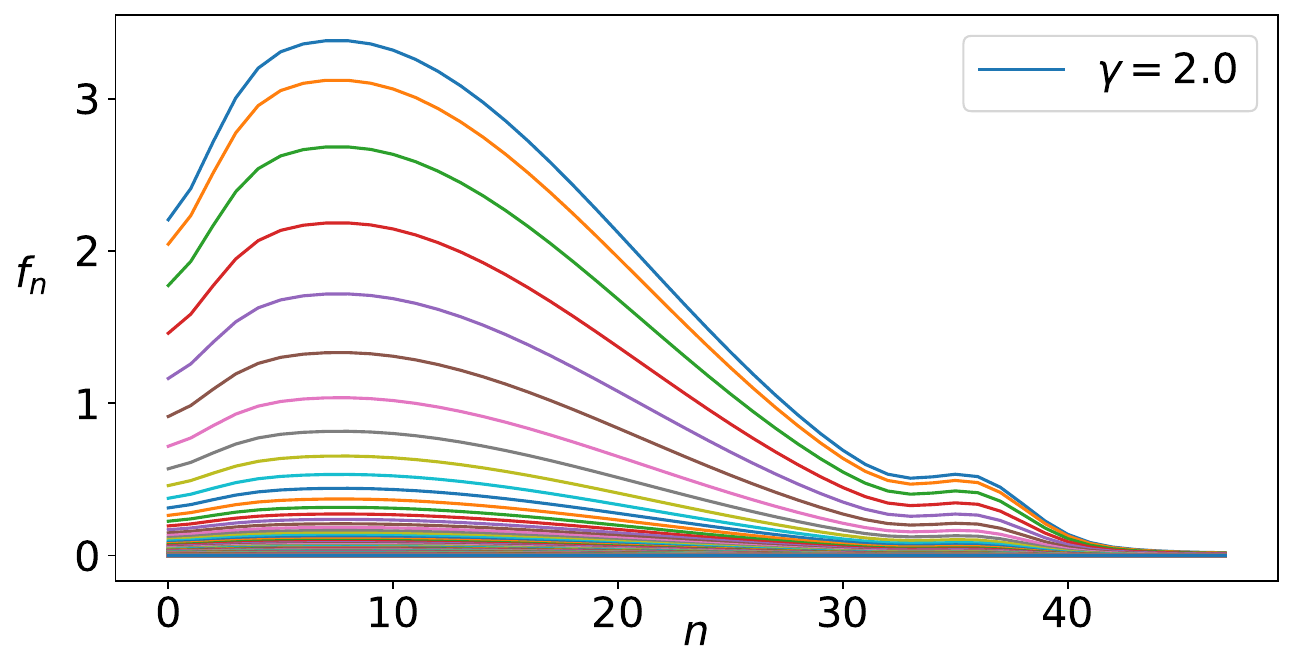}}
    \hspace{0.04\linewidth}
    \caption{$f_{n}$ vs $n$ for a non-Hermitian $\textrm{SSH}$ lattice model with 20 unit cells for three different $\gamma$ values\,. Here we have done the numerical calculation with a time step of 0.1 and shown the results for each time-step.}
    \label{Fig-11}
\end{figure}
\textcolor{black}{Before closing this section we make some comments on whether we can interpret  $F_{Q}^{\textrm{diag}}(\rho(t),\hat{O})$ defined in \eqref{QFIop2}, as generalized complexity functional. To do so, we investigate the behaviour of $f_n$ which follows from \eqref{QFIop2} with respect to $n.$ Following the discussion of \cite{Balasubramanian:2022tpr}, to interpret $F_{Q}^{\textrm{diag}}(\rho(t),\hat{O})$ as a generalized complexity functional,  we will be requiring the $f_{n}'s$ to be positive as well as monotonically increasing function of $n$. Now from the Fig.~(\ref{Fig-11}), we can see that, $f_n$  is always positive but it attains a peak for certain values of $n$ and then decreases before finally saturating to a small value. Since $f_{n}$ is not satisfying the monotonicity  property for our case, so we perhaps cannot interpret it as a generalised measure of spread complexity (in a strict complexity-theoretic senses). Hence, we have claimed it to be structurally similar to the Krylov complexity functional in all of our above discussions.}

\section{Discussion and outlook} \label{section-7}

Motivated by the recent studies of the dynamics of quantum many-body systems in Krylov space and the complexity associated with the spread of state/operator in the Krylov space \cite{PhysRevX.9.041017,Nandy:2024htc}, we have initiated a study of the dynamics of a monitored SSH model (in no-click limit) in Krylov space. This model not only shows a $\mathrm{PT}$ transition but also gives rise to a measurement-induced entanglement phase in the $\mathrm{PT}$ broken phase \cite{10.21468/SciPostPhys.14.5.138}. We first observed that, for this non-Hermitian $\mathrm{SSH}$ chain, measures like \textit{spread complexity, spread entropy, and entropic complexity show oscillatory behaviour below the $\textrm{PT}$ symmetric point. On the other hand, they show a linear behaviour at early times followed by a late time saturation after the $\textrm{PT}$ symmetric point}. It persists even across the measurement-induced transition point, i.e., across $\gamma=2\,$ This is consistent with the earlier observations in \cite{Bhattacharya:2024hto} for a tight-binding chain with non-Hermitian potential at the two edges \footnote{In \cite{Sahu:2024urf}, similar conclusions regarding the insensitivity of the spread complexity for the full system towards the measurement rates have been found.}. \textcolor{black}{It might be interesting to study the possibility of a localization-delocalization transition (in real space) and the effect of edge mode on it for our model and whether it can be detected through the spread complexity (or associated measure in Krylov space), which we leave  for a future study}. \par   Next, we have found that in the $\mathrm{PT}$ broken region, the saturation value of these measures 
decreases as the value of $\gamma$ increases. With the increase in the value of $\gamma$, the wave function becomes more localized, and due to this reason, the suppression of spread complexity and the other associated measures happens. We also found that the saturation timescale ($t_{\mathrm{sat}}$) is an increasing function of $\gamma\,.$ Furthermore, keeping in mind the entanglement transition, we have investigated \textit{the scaling of the spread complexity with system size at a late time, and it takes the following form: $C_{k}\propto L^{\alpha},$ where the scaling exponent $\alpha$ smoothly decreases with the increasing value of $\gamma$ irrespective of the entanglement transition.} Overall, although the late time behaviour of the spread complexity (for the full system) shows different behaviour across the PT transition, it fails to capture the entanglement transition.\par 

Next, we extended the notion of the spread complexity for a subsystem in the Krylov space by utilizing purification techniques.  \textit{We found that scaling of the Krylov complexity of purification ($\textrm{kCoP}$) with the sub-system size. 
The scaling exponent displays a sharp minimum at the PT transition point, i.e., $\gamma=1$, but keeps increasing beyond that irrespective of entanglement transition.} Also, we have computed the time-averaged $\textrm{kCoP}$ and find that it has a dip near $\textrm{PT}$ transition point. This behaviour is more prominent for larger sub-system sizes.  \textcolor{black}{It will be good to get further insight into the physical origin of this behaviour.}  Overall, \textit{we can see that the $\textrm{kCoP}$  displays a sharp transition across the PT transition point and can serve as a good indicator (better than the spread complexity for the full system) of probing such transition} but like the full system complexity it again fails to capture the entanglement transition. 

\par Unlike the entanglement entropy, the spread complexity functional is insensitive to the correlation present in the underlying system. But in recent times, in \cite{Shi:2024bpu, PhysRevLett.133.110201}, it has been shown that the \textit{Quantum Fisher Information} (QFI) when evaluated in the Krylov space, possesses structural similarity to the Krylov complexity but contains information about the correlation landscape of the underlying system. In light of this, we explore this quantity to see whether it can detect the entanglement transition. We calculate $\textrm{QFI}$ for the measurement operator by expanding either the state or the operator in the Krylov space. \textit{From both these two cases, we have found that $\textrm{QFI}$ (in Krylov space) decreases with increasing value of $\gamma$, changes its slope at $\gamma=1$ and almost becomes independent of $\gamma$ around $\gamma=2$.} From this behaviour of $\textrm{QFI}$, we can conclude that $\textrm{QFI}$ can be a promising \textcolor{black}{probe of the  measurement-induced phase} along with a good indicator of the $\textrm{PT}$ transition. \textit{We have further constructed a generalized measure in the Krylov space from the diagonal part of the QFI, and it also shows the same behaviour.} \textcolor{black}{Although the time-averaged QFI (or its diagonal part) displays different behavior in the two phases across $\gamma=2$, we are unable to pinpoint the exact transition point as the slop of it starts to saturate a bit before this value and then stays very low inside the measurement-induced `area law' phase compared to the `volume law' regime (at $\gamma < 2$) regardless of the sub-system size. We leave it for future studies to pinpoint this transition point \footnote{\textcolor{black}{Perhaps one can try to extend our results for even larger (sub) system sizes to see whether the saturation of the QFI happens exactly at $\gamma=2\,.$}}}. \par 
\textcolor{black}{The behaviour of QFI depends on the underlying system as well as the choices of measurement operators, which is evident from the studies of \cite{Poggi:2023mll, Lira-Solanilla:2024alp} for other scenarios. Our results presented in this paper further add to these studies by observing that time-averaged QFI demonstrates distinct behaviour across the above-mentioned two transitions (which are also marked by the change in the nature of the energy spectrum of the underlying model). It further demands more study; in particular, an extension of the analysis of \cite{Lira-Solanilla:2024alp} for the model considered in the paper will be helpful to get further physical insights. We hope to do that in the future, as well as the idea presented in this paper of using $\textrm{QFI}$ in the Krylov space to probe measurement-induced phases needs to be investigated for various other interesting monitored models. We hope to report on it in the near future. }
\par
Finally, we conclude by pointing out some more future directions. First of all, it will be interesting to generalize our study beyond the no-click limit. For that, the time evolution will be governed by the full stochastic Schr\"{o}dinger equation. Then, one can perhaps find the Lanczos coefficients using the moment method from the overlap of the final and initial state following \cite{viswanath1994recursion,Balasubramanian:2022tpr,Caputa:2024vrn}. Furthermore, it will be interesting to generalize the computation of spread complexity for a subsystem in the position space for a time-evolved state. This will perhaps help us to make a more direct comparison with the result of entanglement entropy. Again, one will have to use the idea of purification. We hope to report on some of these in the near future. 

\section*{Acknowledgment}
AB would like to thank the Department of Physics of BITS Pilani, Goa Campus, and Ashoka University for their hospitality during the course of this work. AB is supported by the Core Research Grant (CRG/2023/ 001120) by the Department of Science and Technology Science and Engineering Research Board (India), India. NN is supported by the CSIR fellowship provided by the Govt. of India under the CSIR-JRF scheme (file no. 09/1031(19779)/2024-EMR-I). NC is supported by the Director's Fellowship of the Indian Institute of Technology Gandhinagar. AB also acknowledges the associateship program of the Indian Academy of Sciences (IASc), Bengaluru, as well as the audience of the 90th Annual meeting of IASc at NISER, Bhubaneswar, where part of the work was presented. 

\newpage
\appendix
\section{Bi-Lanczos algorithm  overview}\label{app-A}
Krylov operator/state complexity is usually defined by the well-known Lanczos coefficients \cite{PhysRevX.9.041017,viswanath1994recursion}. However, this is valid only for the case where the time evolution is unitary, i.e., the Hamiltonian is Hermitian. Modifications in the Lanczos algorithm are required to define the Krylov operator/state complexity for non-unitary evolution. Arnoldi's recursive technique has been applied to extend the Lanczos algorithm for such non-Hermitian cases \cite{Bhattacharya:2022gbz,Bhattacharyya:2023grv}. However, this technique converts the Hamiltonian into an upper-Hessenberg matrix instead of a tri-diagonal form. The bi-Lanczos algorithm can deal with this problem, which gives us a tri-diagonal form of a non-Hermitian matrix \cite{Bhattacharya:2023zqt, Bhattacharya:2023yec}. For our studies presented in this paper, we will be required to expand both a state and an operator in terms of the basis generated by the bi-Lanczos algorithm. We discuss them separately below.
\subsection*{For states:}
The non-Hermitian matrix acts differently on the ket vector and bra vector. Therefore, instead of defining a single initial state, we introduce two initial vectors/states $\ket{r} $ and $\bra{l}= \ket{r}^{\dag}$. Using the two-sided Gram-Schmidt procedure for $\ket{r_1}$ and $\bra{l_1}$, we get the Krylov subspaces $\{A\ket{r_1},A^2\ket{r_1},\ldots\}$ and $\{\bra{l_1}A,\bra{l_1}A^2,\ldots\}$ spanned by $\{\ket{r_1}\}$ and $\{\bra{l_1}\}$, respectively. Using the two initial vectors, further vectors are constructed (following bi-orthogonality with each iteration) using the following recursion relations:
\begin{align*}
  c_{j+1}\ket{r_{j+1}}= A\ket{r_j}-a_j\ket{r_j}-b_j\ket{r_{j-1}}\\
b^*_{j+1}\ket{l_{j+1}}= A^\dagger \ket{l_{j}}-a^*_j\ket{l_j}-c^*_j\ket{l_{j-1}}.  
\end{align*}

The non-Hermitian matrix $A$ takes the following tri-diagonal form in this new basis:\\
$$\begin{bmatrix}
    a_1 & b_2 & 0 & \ldots & 0\\
    c_2 & a_2 & b_3 &  & 0 \\
    0 & \ddots & \ddots & \ddots &  \\
    \vdots &  & c_{j-1} & a_{j-1} & b_{j}\\
    0 & \ldots & 0 & c_j & a_j\\
\end{bmatrix}.$$
Here, we present a step-by-step bi-Lanczos algorithm for a non-Hermitian matrix A, which provides us with all the coefficients required to define the Krylov complexity.\\
\newline
For $\textbf{j=1}$:\\\\
(a) Choose the initial set of vectors $\ket{r_1}$ and $\ket{l_1}$ with orthonormality condition $\braket{l_1}{r_1}=1$ and initial lanczos coefficients as $a_1=\bra{l_1}A\ket{r_1}$, $b_1=0$ and $c_1=0$.\\\\
For $\textbf{j>1}$:\\\\\
(a) Act matrix A on $\ket{r_j}$ and $A^{\dag}$ on $\ket{l_j}$. We obtain the following new vectors:\\
\begin{equation}
\ket{\alpha_j'}= A\ket{r_j}\,,\quad
\ket{\beta_j'}= A^{\dagger}\ket{l_j}.
\end{equation}
(b) Construct orthogonal basis vectors by subtracting the contribution from two previous basis vectors $\ket{r_j}$, $\ket{r_{j-1}}$ and similarly for $\ket{l_j}$ and $\ket{l_{j-1}}$.
The new orthogonal vectors are:\\
\begin{equation}
\ket{\alpha_j}= \ket{\alpha_j'}-a_j\ket{r_j}-b_j\ket{r_{j-1}}
\,,\quad 
\ket{\beta_j}= \ket{\beta_j'}-a^*_j\ket{l_j}-c^*_j\ket{l_{j-1}}.    
\end{equation}
\\
(c) The value of the next Lanczos coefficients $c_{j+1}$ and $b_{j+1}$ is given by:
\begin{equation}
c_{j+1}=\sqrt{\abs{\omega_j}}\,,\quad 
b_{j+1}=\frac{\omega_j}{c_{j+1}}    
\end{equation}\\
where, $\omega_j$ is obtained by taking the inner product of the vectors $\ket{\alpha_j}$ and $\ket{\beta_j}$ : $\omega_j = \langle{\alpha_j}|{\beta_j}\rangle$.\\ \\
(d) Since $\ket{\alpha_j}$ and $\ket{\beta_j}$ are not normalised, construct normalized basis vectors $\ket{\alpha_{j+1}}$ and $\ket{\beta_{j+1}}$ of Krylov basis:
\begin{equation}
\ket{r_{j+1}} = \frac{\ket{\alpha_j}}{c_{j+1}}\,,\quad
\ket{l_{j+1}} = \frac{\ket{\beta_j}}{b^*_{j+1}}\,.    
\end{equation}
\\
(e) For successive iterations, subtracting the contribution from only the previous two basis vectors can lead to a significant numerical error. It can be resolved by the Gram-Schmidt orthogonalization technique, which subtracts the contribution of all the previous vectors.\par
For the bi-Lanczos algorithm, bi-orthogonalization is used:
\begin{equation}
\ket{r_j}=\ket{r_j}- \sum_{m=1}^{j-1}\langle{l_m}|{r_m}\rangle\ket{r_m} \,,  
\end{equation}
\begin{equation}
\ket{l_j}=\ket{l_j}- \sum_{m=1}^{j-1}\langle{r_m}|{l_m}\rangle\ket{l_j}\,.
\end{equation}
\\
(f) Finally, the diagonal element of the tridiagonal matrix $a_{j+1}$ is obtained by:
\begin{equation}
a_{j+1}=\bra{l_{j+1}}A\ket{r_{j+1}}\,. 
\end{equation}
\subsection*{For operators:}

The operator equation for non-Hermitian evolution is given by,
\begin{equation}
    \hat{O}(t)= e^{iH^{\dag}t}\hat{O}(0)e^{-iHt}\,.
\end{equation}
The equation of motion satisfied by the operator $\hat{O}(t)$,
\begin{equation}
    \partial_{t}\hat{O}(t)= H^{\dag}\hat{O}(t)-\hat{O}(t)H\,.
\end{equation}
Then, the Liouvillian super-operator can be defined by,
\begin{equation}
    \mathcal{L}\hat{O}(t)=H^{\dag}\hat{O}(t)-\hat{O}(t)H\,.
\end{equation}
Now, we will discuss the bi-Lanczos algorithm step by step.\\\\\\
For $\textbf{j=1}$:\\\\
(a) Choose the initial set of operators $\hat{O}_{r_1}$ and $\hat{O}_{l_1}$ with orthonormality condition \\
$\langle\hat{O}_{l_1}|\hat{O}_{r_1}\rangle=1$, where the inner product of two operators is defined as,
\begin{equation}
   \langle{\hat{O}_{1}}|{\hat{O}_{2}}\rangle=\frac{1}{\sqrt{D}}\Tr(\hat{O}_{1}^{\dag}\hat{O}_{2})
\end{equation} 
and the initial Lanczos coefficients are $b^{1}_{\textrm{up}}=0$ and $b^{1}_{\textrm{down}}=0\,.$\\\\\\
For $\textbf{j>1}$:\\\\
(a) Acting with the Liouvillain super-operator on $\hat{O}_{r_j}$ and $\hat{O}_{l_j}$, we can obtain the new operators.
\begin{equation}
\hat{O}_{\alpha_j'}= H^{\dag}\hat{O}_{r_j}- \hat{O}_{r_j} H
,\hspace{0.2cm}
\hat{O}_{\beta_j'}= H^{\dag}\hat{O}_{l_j}- \hat{O}_{l_j} H\,.
\end{equation}
(b) Construct orthogonal basis vectors by subtracting the contribution from two previous basis operators $\hat{O}_{r_j}$, $\hat{O}_{r_{j-1}}$ and similarly for $\hat{O}_{l_j}$ and $\hat{O}_{l_{j-1}}$.
The new orthogonal vectors are:\\
\begin{equation}
\hat{O}_{\alpha_j}= \hat{O}_{\alpha_j'}-b^{\textrm{up}}_j\hat{O}_{r_{j-1}}\,,\quad
\hat{O}_{\beta_j}= \hat{O}_{\beta_j'}-(b^{\textrm{down}}_j)^{*}\,\hat{O}_{l_{j-1}}\,.   
\end{equation}
\\
(c) The value of the next Lanczos coefficients $b^{\textrm{up}}_{j+1}$ and $b^{\textrm{down}}_{j+1}$ is given by:
\begin{equation}
b^{\textrm{down}}_{j+1}=\sqrt{\abs{\omega_j}}\,,\quad
b^{\textrm{up}}_{j+1}=\frac{\omega_j}{b^{\textrm{down}}_{j+1}}\,.    
\end{equation}\\
where $\omega_j=\langle{\hat{O}_{\alpha_{j}}}|\hat{O}_{\beta_{j}}\rangle\,.$\\ \\
(d) We can then construct the normalized basis vectors $\hat{O}_{\alpha_{j+1}}$ and $\hat{O}_{\beta_{j+1}}$ :
\begin{equation}
\hat{O}_{r_{j+1}} = \frac{\hat{O}_{\alpha_j}}{b^{\textrm{down}}_{j+1}}\,,\quad
\hat{O}_{l_{j+1}} = \frac{\hat{O}_{\beta_j}}{(b^{\textrm{up}}_{j+1})^{*}}\,.
\end{equation}
\\
(e) For successive iterations, subtracting the contribution from only the previous two basis vectors can lead to significant numerical error. It can be resolved by the Gram-Schmidt orthogonalization technique, which subtracts the contribution of all the previous vectors.
\begin{align}
\begin{split}
&\hat{O}_{r_j}=\hat{O}_{r_j}- \sum_{m=1}^{j-1}\langle{\hat{O}_{l_{m}}}|{\hat{O}_{r_{m}}}\rangle \hat{O}_{r_{m}}\,,\\&    
\hat{O}_{l_j}=\hat{O}_{l_j}- \sum_{m=1}^{j-1}\langle{\hat{O}_{r_{m}}}|{\hat{O}_{l_{m}}}\rangle \hat{O}_{l_{m}}\,.      
\end{split} \label{Appeq1}
\end{align}
Finally, when $\omega_{j}=0$ we stop the iteration; otherwise, we have to repeat the calculations from $(a)$ to $(e)$.

\section{Normalisation of wave function }\label{app-B}
As we evolve the state with time, it is important to keep track of normalization, since the total probability must be unity. For Hermitian systems, the evolution is unitary; hence, the evolved state is normalized at every point in time. If we consider the case of effective non-Hermitian Hamiltonian $H=H_1 + iH_2$, where $H_1$ and $H_2$ are Hermitian, the time evolution of the state is done in the following way:
\begin{equation}
 \ket{\psi(t)}  =  e^{-i Ht} \ket{\psi(0)}
=  e^{-i H_1t}e^{H_2t} \ket{\psi(0)}.
\end{equation}
The factor $\exp(H_2t)$ causes a change in the normalisation of the state. One has to correct it to maintain normalisation unity at all later times.
We start with how the density matrix evolves with time when the Hamiltonian is non-Hermitian.
\begin{equation}
    \rho(t) =  \frac{e^{-iHt}\rho (0) e^{i H^{\dag}t}}{\Tr[e^{-iHt}\rho (0) e^{i H^{\dag}t}]}\,.
\end{equation}
Here, $\Tr[e^{-i Ht}\rho (0) e^{i H^{\dag}t}]$ keeps the normalisation to unity even for non-unitary evolution. In Krylov basis, we write initial state vector $\ket{\phi(0)}$ with dimension $k$ which is same as the vector $\ket{\psi_n(0)} = \delta_{n,0}$ 
\begin{equation}
    \ket{\phi(0)} = \begin{bmatrix}
        1\\
        0\\
        0\\
        \vdots\\
        0\\
    \end{bmatrix}\,.
\end{equation}
The density matrix for this state is:
\begin{equation}
\rho(0) = \ket{\phi(0)}\bra{\phi(0)}\,.
\end{equation}
In matrix form, 
\begin{equation}
\rho(0) = \begin{bmatrix}
        1 & 0 & 0 & \ldots & 0 \\
        0 & 0 &   & \\
        0 &  & 0 &  \\
        \vdots & & & \ddots\\
        0 & & &\\
    \end{bmatrix}\,.
\end{equation}
The time-evolved density matrix in the Krylov space can be written in the same manner as before. Here, we use the tri-diagonalized matrix instead of the original non-Hermitian, which is obtained by the bi-Lanczos algorithm to evolve it.
\begin{equation}
    \rho(t) =  \frac{e^{-iLt}\rho (0) e^{i L^{\dag}t}}{\Tr[e^{-i Lt}\rho (0) e^{iL^{\dag}t}]}\,.
\end{equation}
In terms of evolved states, the density matrix is:
\begin{equation}
    \rho(t)= |\tilde{\phi}(t)\rangle\langle\tilde{\phi}(t)|\,.
\end{equation}
 From this, we can define the time-evolved normalized states as: 
 \begin{equation}\label{krylov space rho}
     |\tilde{\phi}(t)\rangle = \frac{e^{-iLt}\ket{\phi(0)}}{\sqrt{\Tr[e^{-i Lt}\rho (0) e^{i L^{\dag}t}]}}\,,\quad \textrm{and} \quad \Big|\langle\tilde{\phi}(t)|\tilde{\phi}(t)\rangle\Big| = 1\,.
 \end{equation}
    
    \section{Dynamics of spread complexity for non-Hermitian \textrm{SSH} model with periodic boundary condition}\label{app-d}
\textcolor{black}{In this appendix, we show the results for the dynamics of the complexity of the non-Hermitian $\textrm{SSH}$ model with periodic boundary conditions. We have performed the numerical computation for 20 unit cells starting with the same initial state (i.e, localized state at the 15th site) as the open boundary case, and the values of all other parameters are fixed to the same value as the open boundary case. We can see from Fig.~(\ref{Fig-13}) that spread complexity, spread entropy, and entropic complexity oscillates below $\gamma=1$, no of oscillations decreases around the $\textrm{PT}$ transition point $\gamma=1$. Beyond $\gamma=1$, these quantities initially rise and then saturate. Hence, we can easily conclude that the main features of these quantities remain the same even if we use the periodic boundary condition instead of the open boundary condition. 
\begin{figure}[htb!]
    \centering
     \captionsetup{font=small}
    \subcaptionbox{$\textrm{Spread Complexity}$}[0.40\linewidth]{\includegraphics[width=1.1\linewidth]{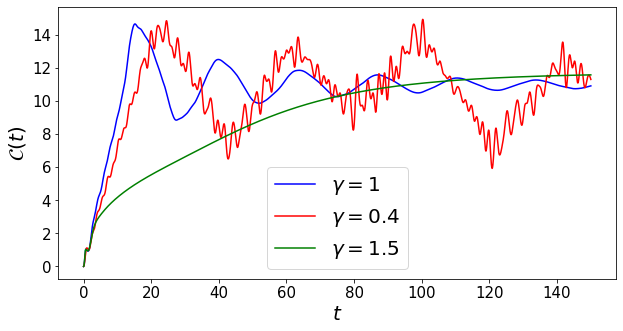}}  \label{Fig-13a}
    \hspace{0.05\linewidth}
    \subcaptionbox{$\textrm{Spread Entropy}$}[0.40\linewidth]{\includegraphics[width=1.1\linewidth]{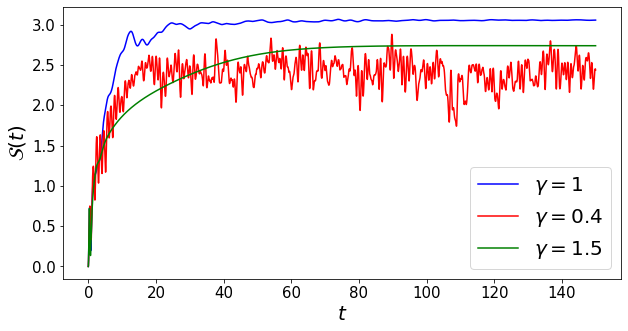}}
    \hspace{0.05\linewidth}
    \subcaptionbox{$\textrm{Entropic complexity}$}[0.40\linewidth]{\includegraphics[width=1.1\linewidth]{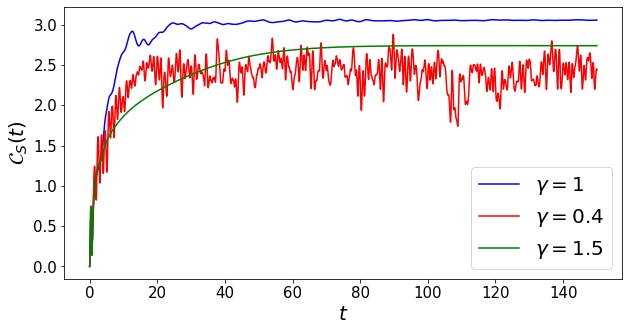}}
    \hspace{0.05\linewidth}
    \caption{Time evolution of spread complexity, spread entropy, and entropic complexity for a non-Hermitian SSH lattice model with 20 unit cells for periodic boundary condition (initial state localized at 15th site)\,.}
    \label{Fig-13}
\end{figure}}
\newpage
\textcolor{black}{Furthermore, we have considered a different initial state localized at the 39th site of the lattice and repeated the numerical calculation to study complexity dynamics. As we can see from Fig.~(\ref{Fig-14}), the overall features of complexity dynamics remain the same.
\begin{figure}[H]
    \centering
    \subcaptionbox{$\textrm{Spread Complexity}$}[0.35\linewidth]{\includegraphics[width=1.1\linewidth]{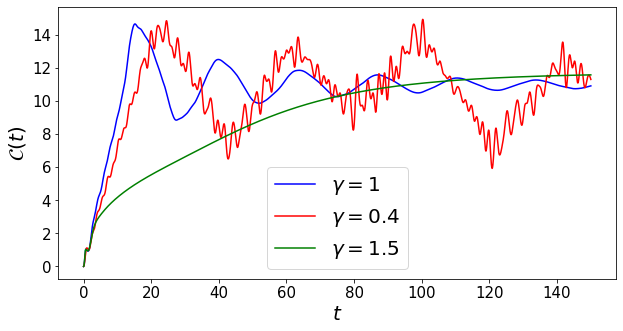}}  \label{Fig-14a}
    \hspace{0.05\linewidth}
    \subcaptionbox{$\textrm{Spread Entropy}$}[0.35\linewidth]{\includegraphics[width=1.1\linewidth]{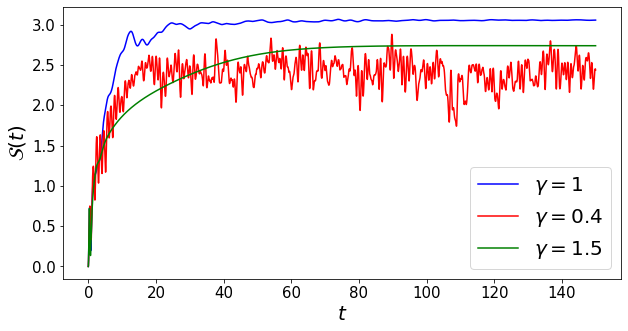}}
    \hspace{0.05\linewidth}
    \subcaptionbox{$\textrm{Entropic complexity}$}[0.35\linewidth]{\includegraphics[width=1.1\linewidth]{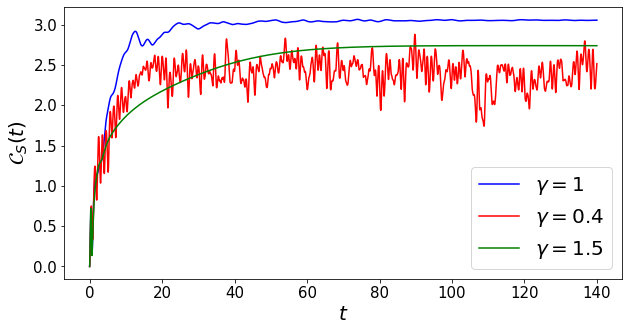}}
    \hspace{0.05\linewidth}
    \caption{Time evolution of spread complexity, spread entropy, and entropic complexity for a non-Hermitian SSH lattice model with 20 unit cells for periodic boundary condition (initial state localized at 39th site)\,.}
    \label{Fig-14}
\end{figure}}
\section{Localization in Krylov space for various choice of initial states}\label{app-c}
\textcolor{black}{We now study the localization in Krylov space for various choice of initial states. We have changed the initial state, keeping all the simulation parameters the same as those mentioned in the main text below \eqref{KIPR eqn}. From Fig.~(\ref{in2}), it is evident that the main features of the dynamics of complexity remain the same even if we change the location of the single particle in the initial state. 
\begin{figure}[htb!]
    \centering
\includegraphics[width=0.45\linewidth]{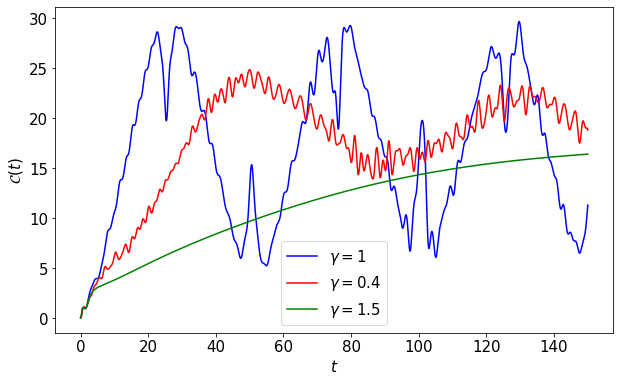}
    \caption{Time evolution of spread complexity for a non-Hermitian SSH model with $20$ unit cells above and below the $\textrm{PT}$ transition (initial state localized at the 5th site)\,.}
    \label{in2}
\end{figure}}
\par
\textcolor{black}{Furthermore, we have chosen three different initial states, where the particle is uniformly delocalized over two sites, keeping the system size and other simulation parameters the same as before. Again, we have checked that the qualitative behaviour of the spread complexity below and above the PT transition point (i.e, $\gamma=1$) remains the same. We focus mainly on the saturation value of the spread complexity (above $\gamma=1$) below. We have checked that this nature persists even beyond $\gamma=2$ i.e., the entanglement transition point.
\begin{figure}[htb!]
    \centering   
 \includegraphics[width=0.50\linewidth]{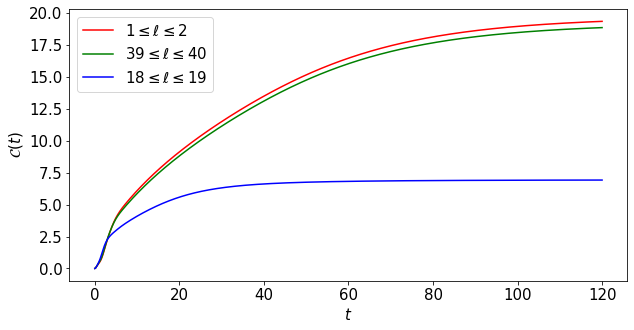} 
    \caption{Spread Complexity in the \textrm{PT} broken phase (for $\gamma=1.2$) for three different initial states.}
    \label{Initial state dependence}
\end{figure}}
\par
\textcolor{black}{We can see from Fig.~(\ref{Initial state dependence}) that if there is an initial spread (in the position space) near the two edges of the lattice chain, then the spread complexity (saturation value of it)  is greater, and if the initial spread (in the position space) is over some sites around the middle of the chain, then the saturation value of the spread complexity is less which can also be understood from the value $\textrm{KIPR}$ \footnote{\textcolor{black}{One can check that the time-evolved steady state in the lattice site basis when the initial wave-packet is delocalized over the two central sites is itself symmetric around the chain centre compared to the case when the initial wave-packet is over the edges. Hence, little additional spreading in Krylov space is required for the former case, indicating a smaller value of the spread complexity at later times.}}\,. From Fig.~(\ref{KIPR initial state dependence}), we can see that the saturation value of $\textrm{KIPR}$ is greater when the initial spread is near the middle of the chain, which implies higher localization in the Krylov space and hence less amount of spread complexity.
\begin{figure}[H]
    \centering  \includegraphics[width=0.50\linewidth]{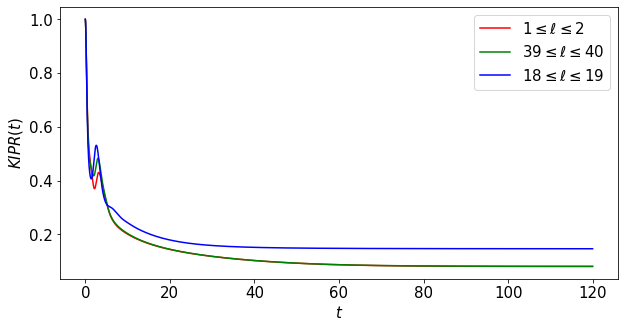}
    \caption{KIPR in the PT broken phase (for $\gamma=1.2$) for three different initial states.}
    \label{KIPR initial state dependence}
\end{figure}}

\section{\textrm{QFI} dynamics in Krylov space with periodic boundary condition }\label{app-e}

\textcolor{black}{We show how time-averaged $\textrm{QFI}$ changes with $\gamma$ for a non-Hermitian SSH model for $L=16,20,24$ unit cells with periodic boundary conditions. We can see from the Fig.~(\ref{QFIPB}) that time-averaged $\textrm{QFI}$ ($\bar{F}_{Q}$) decreases with increasing value of $\gamma$, changes slope near $\gamma=1$ and almost becomes independent of $\gamma$ near $\gamma=2$ which is shown in the inset. So, the behaviour of the $\bar{F}_{Q}$ remains the same (like the open boundary case discussed in the main text) in the presence of boundary conditions.
\begin{figure}[htb!]
    \centering
    \includegraphics[width=0.55\linewidth]{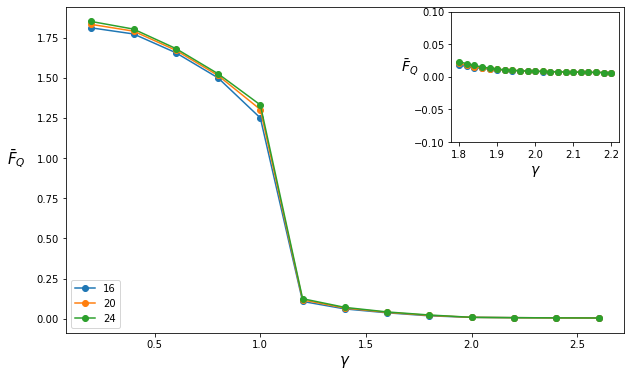}
    \caption{Time averaged \textrm{QFI} with $\gamma$ for different system sizes. In the inset we have shown the behaviour of $\textrm{QFI}$ near $\gamma=2\,.$} 
    \label{QFIPB}
\end{figure}}



 

\bibliographystyle{JHEP}
\bibliography{SSHbibliography}

\providecommand{\href}[2]{#2}\begingroup\raggedright\begin{thebibliography}{100}

\bibitem{NL1}
M.A.~Nielsen, \emph{{A geometric approach to quantum circuit lower bounds}},
  \href{https://doi.org/10.3390/universe5040092}{\emph{Science} {\bfseries 311}
  (2006) 92} [\href{https://arxiv.org/abs/0502070}{{\ttfamily 0502070}}].

\bibitem{NL2}
M.A.~Nielsen, M.R.~Dowling, M.~Gu and A.C.~Doherty, \emph{Quantum computation
  as geometry}, \href{https://doi.org/10.1126/science.1121541}{\emph{Science}
  {\bfseries 311} (2006) 1133}
  [\href{https://arxiv.org/abs/https://www.science.org/doi/pdf/10.1126/science.1121541}{{\ttfamily
  https://www.science.org/doi/pdf/10.1126/science.1121541}}].

\bibitem{NL3}
M.R.~Nielsen, M.~A.and~Dowling, \emph{{The geometry of quantum computation}},
  \href{https://doi.org/10.3390/universe5040092}{\emph{Science} {\bfseries 311}
  (2006) 1133} [\href{https://arxiv.org/abs/0701004}{{\ttfamily 0701004}}].

\bibitem{Jefferson}
R.~Jefferson and R.C.~Myers, \emph{{Circuit complexity in quantum field
  theory}}, \href{https://doi.org/10.1007/JHEP10(2017)107}{\emph{JHEP}
  {\bfseries 10} (2017) 107}
  [\href{https://arxiv.org/abs/1707.08570}{{\ttfamily 1707.08570}}].

\bibitem{Chapman:2017rqy}
S.~Chapman, M.P.~Heller, H.~Marrochio and F.~Pastawski, \emph{{Toward a
  Definition of Complexity for Quantum Field Theory States}},
  \href{https://doi.org/10.1103/PhysRevLett.120.121602}{\emph{Phys. Rev. Lett.}
  {\bfseries 120} (2018) 121602}
  [\href{https://arxiv.org/abs/1707.08582}{{\ttfamily 1707.08582}}].

\bibitem{Bhattacharyya:2018wym}
A.~Bhattacharyya, P.~Caputa, S.R.~Das, N.~Kundu, M.~Miyaji and T.~Takayanagi,
  \emph{{Path-Integral Complexity for Perturbed CFTs}},
  \href{https://doi.org/10.1007/JHEP07(2018)086}{\emph{JHEP} {\bfseries 07}
  (2018) 086} [\href{https://arxiv.org/abs/1804.01999}{{\ttfamily
  1804.01999}}].

\bibitem{Caputa:2017yrh}
P.~Caputa, N.~Kundu, M.~Miyaji, T.~Takayanagi and K.~Watanabe, \emph{{Liouville
  Action as Path-Integral Complexity: From Continuous Tensor Networks to
  AdS/CFT}}, \href{https://doi.org/10.1007/JHEP11(2017)097}{\emph{JHEP}
  {\bfseries 11} (2017) 097}
  [\href{https://arxiv.org/abs/1706.07056}{{\ttfamily 1706.07056}}].

\bibitem{Ali:2018fcz}
T.~Ali, A.~Bhattacharyya, S.~Shajidul~Haque, E.H.~Kim and N.~Moynihan,
  \emph{{Time Evolution of Complexity: A Critique of Three Methods}},
  \href{https://doi.org/10.1007/JHEP04(2019)087}{\emph{JHEP} {\bfseries 04}
  (2019) 087} [\href{https://arxiv.org/abs/1810.02734}{{\ttfamily
  1810.02734}}].

\bibitem{Bhattacharyya:2018bbv}
A.~Bhattacharyya, A.~Shekar and A.~Sinha, \emph{{Circuit complexity in
  interacting QFTs and RG flows}},
  \href{https://doi.org/10.1007/JHEP10(2018)140}{\emph{JHEP} {\bfseries 10}
  (2018) 140} [\href{https://arxiv.org/abs/1808.03105}{{\ttfamily
  1808.03105}}].

\bibitem{Hackl:2018ptj}
L.~Hackl and R.C.~Myers, \emph{{Circuit complexity for free fermions}},
  \href{https://doi.org/10.1007/JHEP07(2018)139}{\emph{JHEP} {\bfseries 07}
  (2018) 139} [\href{https://arxiv.org/abs/1803.10638}{{\ttfamily
  1803.10638}}].

\bibitem{Khan:2018rzm}
R.~Khan, C.~Krishnan and S.~Sharma, \emph{{Circuit Complexity in Fermionic
  Field Theory}}, \href{https://doi.org/10.1103/PhysRevD.98.126001}{\emph{Phys.
  Rev. D} {\bfseries 98} (2018) 126001}
  [\href{https://arxiv.org/abs/1801.07620}{{\ttfamily 1801.07620}}].

\bibitem{Camargo:2018eof}
H.A.~Camargo, P.~Caputa, D.~Das, M.P.~Heller and R.~Jefferson,
  \emph{{Complexity as a novel probe of quantum quenches: universal scalings
  and purifications}},
  \href{https://doi.org/10.1103/PhysRevLett.122.081601}{\emph{Phys. Rev. Lett.}
  {\bfseries 122} (2019) 081601}
  [\href{https://arxiv.org/abs/1807.07075}{{\ttfamily 1807.07075}}].

\bibitem{Ali:2018aon}
T.~Ali, A.~Bhattacharyya, S.~Shajidul~Haque, E.H.~Kim and N.~Moynihan,
  \emph{{Post-Quench Evolution of Complexity and Entanglement in a Topological
  System}}, \href{https://doi.org/10.1016/j.physletb.2020.135919}{\emph{Phys.
  Lett. B} {\bfseries 811} (2020) 135919}
  [\href{https://arxiv.org/abs/1811.05985}{{\ttfamily 1811.05985}}].

\bibitem{Caputa:2018kdj}
P.~Caputa and J.M.~Magan, \emph{{Quantum Computation as Gravity}},
  \href{https://doi.org/10.1103/PhysRevLett.122.231302}{\emph{Phys. Rev. Lett.}
  {\bfseries 122} (2019) 231302}
  [\href{https://arxiv.org/abs/1807.04422}{{\ttfamily 1807.04422}}].

\bibitem{Guo:2018kzl}
M.~Guo, J.~Hernandez, R.C.~Myers and S.-M.~Ruan, \emph{{Circuit Complexity for
  Coherent States}}, \href{https://doi.org/10.1007/JHEP10(2018)011}{\emph{JHEP}
  {\bfseries 10} (2018) 011}
  [\href{https://arxiv.org/abs/1807.07677}{{\ttfamily 1807.07677}}].

\bibitem{Bhattacharyya:2019kvj}
A.~Bhattacharyya, P.~Nandy and A.~Sinha, \emph{{Renormalized Circuit
  Complexity}},
  \href{https://doi.org/10.1103/PhysRevLett.124.101602}{\emph{Phys. Rev. Lett.}
  {\bfseries 124} (2020) 101602}
  [\href{https://arxiv.org/abs/1907.08223}{{\ttfamily 1907.08223}}].

\bibitem{Erdmenger:2020sup}
J.~Erdmenger, M.~Gerbershagen and A.-L.~Weigel, \emph{{Complexity measures from
  geometric actions on Virasoro and Kac-Moody orbits}},
  \href{https://doi.org/10.1007/JHEP11(2020)003}{\emph{JHEP} {\bfseries 11}
  (2020) 003} [\href{https://arxiv.org/abs/2004.03619}{{\ttfamily
  2004.03619}}].

\bibitem{Ali:2019zcj}
T.~Ali, A.~Bhattacharyya, S.S.~Haque, E.H.~Kim, N.~Moynihan and J.~Murugan,
  \emph{{Chaos and Complexity in Quantum Mechanics}},
  \href{https://doi.org/10.1103/PhysRevD.101.026021}{\emph{Phys. Rev. D}
  {\bfseries 101} (2020) 026021}
  [\href{https://arxiv.org/abs/1905.13534}{{\ttfamily 1905.13534}}].

\bibitem{Bhattacharyya:2019txx}
A.~Bhattacharyya, W.~Chemissany, S.~Shajidul~Haque and B.~Yan, \emph{{Towards
  the web of quantum chaos diagnostics}},
  \href{https://doi.org/10.1140/epjc/s10052-022-10035-3}{\emph{Eur. Phys. J. C}
  {\bfseries 82} (2022) 87} [\href{https://arxiv.org/abs/1909.01894}{{\ttfamily
  1909.01894}}].

\bibitem{Caceres:2019pgf}
E.~Caceres, S.~Chapman, J.D.~Couch, J.P.~Hernandez, R.C.~Myers and S.-M.~Ruan,
  \emph{{Complexity of Mixed States in QFT and Holography}},
  \href{https://doi.org/10.1007/JHEP03(2020)012}{\emph{JHEP} {\bfseries 03}
  (2020) 012} [\href{https://arxiv.org/abs/1909.10557}{{\ttfamily
  1909.10557}}].

\bibitem{Bhattacharyya:2020art}
A.~Bhattacharyya, W.~Chemissany, S.S.~Haque, J.~Murugan and B.~Yan, \emph{{The
  Multi-faceted Inverted Harmonic Oscillator: Chaos and Complexity}},
  \href{https://doi.org/10.21468/SciPostPhysCore.4.1.002}{\emph{SciPost Phys.
  Core} {\bfseries 4} (2021) 002}
  [\href{https://arxiv.org/abs/2007.01232}{{\ttfamily 2007.01232}}].

\bibitem{Liu_2020}
F.~Liu, S.~Whitsitt, J.B.~Curtis, R.~Lundgren, P.~Titum, Z.-C.~Yang et~al.,
  \emph{{Circuit complexity across a topological phase transition}},
  \href{https://doi.org/10.1103/PhysRevResearch.2.013323}{\emph{Phys. Rev.
  Res.} {\bfseries 2} (2020) 013323}
  [\href{https://arxiv.org/abs/1902.10720}{{\ttfamily 1902.10720}}].

\bibitem{Bhattacharyya:2020rpy}
A.~Bhattacharyya, S.~Das, S.~Shajidul~Haque and B.~Underwood,
  \emph{{Cosmological Complexity}},
  \href{https://doi.org/10.1103/PhysRevD.101.106020}{\emph{Phys. Rev. D}
  {\bfseries 101} (2020) 106020}
  [\href{https://arxiv.org/abs/2001.08664}{{\ttfamily 2001.08664}}].

\bibitem{Bhattacharyya:2020kgu}
A.~Bhattacharyya, S.~Das, S.S.~Haque and B.~Underwood, \emph{{Rise of
  cosmological complexity: Saturation of growth and chaos}},
  \href{https://doi.org/10.1103/PhysRevResearch.2.033273}{\emph{Phys. Rev.
  Res.} {\bfseries 2} (2020) 033273}
  [\href{https://arxiv.org/abs/2005.10854}{{\ttfamily 2005.10854}}].

\bibitem{Chen:2020nlj}
B.~Chen, B.~Czech and Z.-z.~Wang, \emph{{Cutoff Dependence and Complexity of
  the CFT$_2$ Ground State}},
  \href{https://arxiv.org/abs/2004.11377}{{\ttfamily 2004.11377}}.

\bibitem{Czech:2017ryf}
B.~Czech, \emph{{Einstein Equations from Varying Complexity}},
  \href{https://doi.org/10.1103/PhysRevLett.120.031601}{\emph{Phys. Rev. Lett.}
  {\bfseries 120} (2018) 031601}
  [\href{https://arxiv.org/abs/1706.00965}{{\ttfamily 1706.00965}}].

\bibitem{Chapman:2018hou}
S.~Chapman, J.~Eisert, L.~Hackl, M.P.~Heller, R.~Jefferson, H.~Marrochio
  et~al., \emph{{Complexity and entanglement for thermofield double states}},
  \href{https://doi.org/10.21468/SciPostPhys.6.3.034}{\emph{SciPost Phys.}
  {\bfseries 6} (2019) 034} [\href{https://arxiv.org/abs/1810.05151}{{\ttfamily
  1810.05151}}].

\bibitem{Couch:2021wsm}
J.~Couch, Y.~Fan and S.~Shashi, \emph{{Circuit Complexity in Topological
  Quantum Field Theory}},  \href{https://arxiv.org/abs/2108.13427}{{\ttfamily
  2108.13427}}.

\bibitem{Chagnet:2021uvi}
N.~Chagnet, S.~Chapman, J.~de~Boer and C.~Zukowski, \emph{{Complexity for
  Conformal Field Theories in General Dimensions}},
  \href{https://doi.org/10.1103/PhysRevLett.128.051601}{\emph{Phys. Rev. Lett.}
  {\bfseries 128} (2022) 051601}
  [\href{https://arxiv.org/abs/2103.06920}{{\ttfamily 2103.06920}}].

\bibitem{Koch:2021tvp}
R.d.M.~Koch, M.~Kim and H.J.R.~Van~Zyl, \emph{{Complexity from spinning
  primaries}}, \href{https://doi.org/10.1007/JHEP12(2021)030}{\emph{JHEP}
  {\bfseries 12} (2021) 030}
  [\href{https://arxiv.org/abs/2108.10669}{{\ttfamily 2108.10669}}].

\bibitem{Bhattacharyya:2022ren}
A.~Bhattacharyya, G.~Katoch and S.R.~Roy, \emph{{Complexity of warped conformal
  field theory}},
  \href{https://doi.org/10.1140/epjc/s10052-023-11212-8}{\emph{Eur. Phys. J. C}
  {\bfseries 83} (2023) 33} [\href{https://arxiv.org/abs/2202.09350}{{\ttfamily
  2202.09350}}].

\bibitem{Bhattacharyya:2023sjr}
A.~Bhattacharyya and P.~Nandi, \emph{{Circuit complexity for Carrollian
  Conformal (BMS) field theories}},
  \href{https://doi.org/10.1007/JHEP07(2023)105}{\emph{JHEP} {\bfseries 07}
  (2023) 105} [\href{https://arxiv.org/abs/2301.12845}{{\ttfamily
  2301.12845}}].

\bibitem{Bhattacharyya:2022rhm}
A.~Bhattacharyya, T.~Hanif, S.S.~Haque and A.~Paul, \emph{{Decoherence,
  entanglement negativity, and circuit complexity for an open quantum system}},
  \href{https://doi.org/10.1103/PhysRevD.107.106007}{\emph{Phys. Rev. D}
  {\bfseries 107} (2023) 106007}
  [\href{https://arxiv.org/abs/2210.09268}{{\ttfamily 2210.09268}}].

\bibitem{Bhattacharyya:2021fii}
A.~Bhattacharyya, T.~Hanif, S.S.~Haque and M.K.~Rahman, \emph{{Complexity for
  an open quantum system}},
  \href{https://doi.org/10.1103/PhysRevD.105.046011}{\emph{Phys. Rev. D}
  {\bfseries 105} (2022) 046011}
  [\href{https://arxiv.org/abs/2112.03955}{{\ttfamily 2112.03955}}].

\bibitem{Bhattacharyya:2020iic}
A.~Bhattacharyya, S.S.~Haque and E.H.~Kim, \emph{{Complexity from the reduced
  density matrix: a new diagnostic for chaos}},
  \href{https://doi.org/10.1007/JHEP10(2021)028}{\emph{JHEP} {\bfseries 10}
  (2021) 028} [\href{https://arxiv.org/abs/2011.04705}{{\ttfamily
  2011.04705}}].

\bibitem{Craps:2023rur}
B.~Craps, M.~De~Clerck, O.~Evnin and P.~Hacker, \emph{{Integrability and
  complexity in quantum spin chains}},
  \href{https://arxiv.org/abs/2305.00037}{{\ttfamily 2305.00037}}.

\bibitem{Jaiswal:2021tnt}
N.~Jaiswal, M.~Gautam and T.~Sarkar, \emph{{Complexity, information geometry,
  and Loschmidt echo near quantum criticality}},
  \href{https://doi.org/10.1088/1742-5468/ac7aa6}{\emph{J. Stat. Mech.}
  {\bfseries 2207} (2022) 073105}
  [\href{https://arxiv.org/abs/2110.02099}{{\ttfamily 2110.02099}}].

\bibitem{Bhattacharya:2022wlp}
A.~Bhattacharya, A.~Bhattacharyya and S.~Maulik, \emph{{Pseudocomplexity of
  purification for free scalar field theories}},
  \href{https://doi.org/10.1103/PhysRevD.106.086010}{\emph{Phys. Rev. D}
  {\bfseries 106} (2022) 086010}
  [\href{https://arxiv.org/abs/2209.00049}{{\ttfamily 2209.00049}}].

\bibitem{Bhattacharyya:2024rzz}
A.~Bhattacharyya, S.~Brahma, S.~Chowdhury and X.~Luo, \emph{{Benchmarking
  quantum chaos from geometric complexity}},
  \href{https://arxiv.org/abs/2410.18754}{{\ttfamily 2410.18754}}.

\bibitem{Chowdhury:2023iwg}
S.~Chowdhury, M.~Bojowald and J.~Mielczarek, \emph{{Geometric quantum
  complexity of bosonic oscillator systems}},
  \href{https://doi.org/10.1007/JHEP10(2024)048}{\emph{JHEP} {\bfseries 10}
  (2024) 048} [\href{https://arxiv.org/abs/2307.13736}{{\ttfamily
  2307.13736}}].

\bibitem{Chapman:2021jbh}
S.~Chapman and G.~Policastro, \emph{{Quantum Computational Complexity -- From
  Quantum Information to Black Holes and Back}},
  \href{https://arxiv.org/abs/2110.14672}{{\ttfamily 2110.14672}}.

\bibitem{Bhattacharyya:2021cwf}
A.~Bhattacharyya, \emph{{Circuit complexity and (some of) its applications}},
  \href{https://doi.org/10.1142/S0218301321300058}{\emph{Int. J. Mod. Phys. E}
  {\bfseries 30} (2021) 2130005}.

\bibitem{Katoch:2023etn}
G.~Katoch, \emph{{Investigations of LST and WCFT using complexity as a probe}},
  Ph.D. thesis, Indian Inst. Tech., Hyderabad, 8, 2023.

\bibitem{Susskind:2014moa}
L.~Susskind, \emph{{Entanglement is not enough}},
  \href{https://doi.org/10.1002/prop.201500095}{\emph{Fortsch. Phys.}
  {\bfseries 64} (2016) 49} [\href{https://arxiv.org/abs/1411.0690}{{\ttfamily
  1411.0690}}].

\bibitem{Brown:2015bva}
A.R.~Brown, D.A.~Roberts, L.~Susskind, B.~Swingle and Y.~Zhao,
  \emph{{Holographic Complexity Equals Bulk Action?}},
  \href{https://doi.org/10.1103/PhysRevLett.116.191301}{\emph{Phys. Rev. Lett.}
  {\bfseries 116} (2016) 191301}
  [\href{https://arxiv.org/abs/1509.07876}{{\ttfamily 1509.07876}}].

\bibitem{Stanford:2014jda}
D.~Stanford and L.~Susskind, \emph{{Complexity and Shock Wave Geometries}},
  \href{https://doi.org/10.1103/PhysRevD.90.126007}{\emph{Phys. Rev. D}
  {\bfseries 90} (2014) 126007}
  [\href{https://arxiv.org/abs/1406.2678}{{\ttfamily 1406.2678}}].

\bibitem{Carmi:2016wjl}
D.~Carmi, R.C.~Myers and P.~Rath, \emph{{Comments on Holographic Complexity}},
  \href{https://doi.org/10.1007/JHEP03(2017)118}{\emph{JHEP} {\bfseries 03}
  (2017) 118} [\href{https://arxiv.org/abs/1612.00433}{{\ttfamily
  1612.00433}}].

\bibitem{PhysRevX.9.041017}
D.E.~Parker, X.~Cao, A.~Avdoshkin, T.~Scaffidi and E.~Altman, \emph{A universal
  operator growth hypothesis},
  \href{https://doi.org/10.1103/PhysRevX.9.041017}{\emph{Phys. Rev. X}
  {\bfseries 9} (2019) 041017}.

\bibitem{Barbon:2019wsy}
J.L.F.~Barb\'on, E.~Rabinovici, R.~Shir and R.~Sinha, \emph{{On The Evolution
  Of Operator Complexity Beyond Scrambling}},
  \href{https://doi.org/10.1007/JHEP10(2019)264}{\emph{JHEP} {\bfseries 10}
  (2019) 264} [\href{https://arxiv.org/abs/1907.05393}{{\ttfamily
  1907.05393}}].

\bibitem{Rabinovici:2020ryf}
E.~Rabinovici, A.~S\'anchez-Garrido, R.~Shir and J.~Sonner, \emph{{Operator
  complexity: a journey to the edge of Krylov space}},
  \href{https://doi.org/10.1007/JHEP06(2021)062}{\emph{JHEP} {\bfseries 06}
  (2021) 062} [\href{https://arxiv.org/abs/2009.01862}{{\ttfamily
  2009.01862}}].

\bibitem{Jian:2020qpp}
S.-K.~Jian, B.~Swingle and Z.-Y.~Xian, \emph{{Complexity growth of operators in
  the SYK model and in JT gravity}},
  \href{https://doi.org/10.1007/JHEP03(2021)014}{\emph{JHEP} {\bfseries 03}
  (2021) 014} [\href{https://arxiv.org/abs/2008.12274}{{\ttfamily
  2008.12274}}].

\bibitem{DymarskyPRB2020}
A.~Dymarsky and A.~Gorsky, \emph{Quantum chaos as delocalization in krylov
  space}, \href{https://doi.org/10.1103/PhysRevB.102.085137}{\emph{Phys. Rev.
  B} {\bfseries 102} (2020) 085137}.

\bibitem{PhysRevLett.124.206803}
D.J.~Yates, A.G.~Abanov and A.~Mitra, \emph{Lifetime of almost strong edge-mode
  operators in one-dimensional, interacting, symmetry protected topological
  phases}, \href{https://doi.org/10.1103/PhysRevLett.124.206803}{\emph{Phys.
  Rev. Lett.} {\bfseries 124} (2020) 206803}.

\bibitem{Yates2020}
D.J.~Yates, A.G.~Abanov and A.~Mitra, \emph{Dynamics of almost strong edge
  modes in spin chains away from integrability},
  \href{https://doi.org/10.1103/physrevb.102.195419}{\emph{Physical Review B}
  {\bfseries 102} (2020) }.

\bibitem{Rabinovici:2021qqt}
E.~Rabinovici, A.~S\'anchez-Garrido, R.~Shir and J.~Sonner, \emph{{Krylov
  localization and suppression of complexity}},
  \href{https://doi.org/10.1007/JHEP03(2022)211}{\emph{JHEP} {\bfseries 03}
  (2022) 211} [\href{https://arxiv.org/abs/2112.12128}{{\ttfamily
  2112.12128}}].

\bibitem{Yates:2021lrt}
D.J.~Yates, A.G.~Abanov and A.~Mitra, \emph{{Long-lived period-doubled edge
  modes of interacting and disorder-free Floquet spin chains}},
  \href{https://arxiv.org/abs/2105.13766}{{\ttfamily 2105.13766}}.

\bibitem{Yates:2021asz}
D.J.~Yates and A.~Mitra, \emph{{Strong and almost strong modes of Floquet spin
  chains in Krylov subspaces}},
  \href{https://doi.org/10.1103/PhysRevB.104.195121}{\emph{Phys. Rev. B}
  {\bfseries 104} (2021) 195121}
  [\href{https://arxiv.org/abs/2105.13246}{{\ttfamily 2105.13246}}].

\bibitem{Dymarsky:2021bjq}
A.~Dymarsky and M.~Smolkin, \emph{{Krylov complexity in conformal field
  theory}}, \href{https://doi.org/10.1103/PhysRevD.104.L081702}{\emph{Phys.
  Rev. D} {\bfseries 104} (2021) L081702}
  [\href{https://arxiv.org/abs/2104.09514}{{\ttfamily 2104.09514}}].

\bibitem{Noh2021}
J.D.~Noh, \emph{Operator growth in the transverse-field ising spin chain with
  integrability-breaking longitudinal field},
  \href{https://doi.org/10.1103/physreve.104.034112}{\emph{Physical Review E}
  {\bfseries 104} (2021) }.

\bibitem{Trigueros:2021rwj}
F.B.~Trigueros and C.-J.~Lin, \emph{{Krylov complexity of many-body
  localization: Operator localization in Krylov basis}},
  \href{https://arxiv.org/abs/2112.04722}{{\ttfamily 2112.04722}}.

\bibitem{Liu:2022god}
C.~Liu, H.~Tang and H.~Zhai, \emph{{Krylov Complexity in Open Quantum
  Systems}},  \href{https://arxiv.org/abs/2207.13603}{{\ttfamily 2207.13603}}.

\bibitem{Fan_2022}
Z.-Y.~Fan, \emph{Universal relation for operator complexity},
  \href{https://doi.org/10.1103/physreva.105.062210}{\emph{Physical Review A}
  {\bfseries 105} (2022) }.

\bibitem{Kar_2022}
A.~Kar, L.~Lamprou, M.~Rozali and J.~Sully, \emph{Random matrix theory for
  complexity growth and black hole interiors},
  \href{https://doi.org/10.1007/jhep01(2022)016}{\emph{Journal of High Energy
  Physics} {\bfseries 2022} (2022) }.

\bibitem{Caputa:2021sib}
P.~Caputa, J.M.~Magan and D.~Patramanis, \emph{{Geometry of Krylov
  complexity}},
  \href{https://doi.org/10.1103/PhysRevResearch.4.013041}{\emph{Phys. Rev.
  Res.} {\bfseries 4} (2022) 013041}
  [\href{https://arxiv.org/abs/2109.03824}{{\ttfamily 2109.03824}}].

\bibitem{PhysRevE.106.014152}
R.~Heveling, J.~Wang and J.~Gemmer, \emph{Numerically probing the universal
  operator growth hypothesis},
  \href{https://doi.org/10.1103/PhysRevE.106.014152}{\emph{Phys. Rev. E}
  {\bfseries 106} (2022) 014152}.

\bibitem{Bhattacharjee_2022a}
B.~Bhattacharjee, X.~Cao, P.~Nandy and T.~Pathak, \emph{Krylov complexity in
  saddle-dominated scrambling},
  \href{https://doi.org/10.1007/jhep05(2022)174}{\emph{Journal of High Energy
  Physics} {\bfseries 2022} (2022) }.

\bibitem{Adhikari:2022whf}
K.~Adhikari, S.~Choudhury and A.~Roy, \emph{{Krylov Complexity in Quantum Field
  Theory}}, \href{https://doi.org/10.1016/j.nuclphysb.2023.116263}{\emph{Nucl.
  Phys. B} {\bfseries 993} (2023) 116263}
  [\href{https://arxiv.org/abs/2204.02250}{{\ttfamily 2204.02250}}].

\bibitem{https://doi.org/10.48550/arxiv.2205.12815}
W.~M\"uck and Y.~Yang, \emph{{Krylov complexity and orthogonal polynomials}},
  \href{https://doi.org/10.1016/j.nuclphysb.2022.115948}{\emph{Nucl. Phys. B}
  {\bfseries 984} (2022) 115948}
  [\href{https://arxiv.org/abs/2205.12815}{{\ttfamily 2205.12815}}].

\bibitem{Bhattacharya:2022gbz}
A.~Bhattacharya, P.~Nandy, P.P.~Nath and H.~Sahu, \emph{{Operator growth and
  Krylov construction in dissipative open quantum systems}},
  \href{https://doi.org/10.1007/JHEP12(2022)081}{\emph{JHEP} {\bfseries 12}
  (2022) 081} [\href{https://arxiv.org/abs/2207.05347}{{\ttfamily
  2207.05347}}].

\bibitem{H_rnedal_2022}
N.~Hörnedal, N.~Carabba, A.S.~Matsoukas-Roubeas and A.~del Campo,
  \emph{Ultimate speed limits to the growth of operator complexity},
  \href{https://doi.org/10.1038/s42005-022-00985-1}{\emph{Communications
  Physics} {\bfseries 5} (2022) }.

\bibitem{Bhattacharjee:2022lzy}
B.~Bhattacharjee, X.~Cao, P.~Nandy and T.~Pathak, \emph{{Operator growth in
  open quantum systems: lessons from the dissipative SYK}},
  \href{https://doi.org/10.1007/JHEP03(2023)054}{\emph{JHEP} {\bfseries 03}
  (2023) 054} [\href{https://arxiv.org/abs/2212.06180}{{\ttfamily
  2212.06180}}].

\bibitem{Rabinovici:2022beu}
E.~Rabinovici, A.~S\'anchez-Garrido, R.~Shir and J.~Sonner, \emph{{Krylov
  complexity from integrability to chaos}},
  \href{https://doi.org/10.1007/JHEP07(2022)151}{\emph{JHEP} {\bfseries 07}
  (2022) 151} [\href{https://arxiv.org/abs/2207.07701}{{\ttfamily
  2207.07701}}].

\bibitem{Alishahiha:2022anw}
M.~Alishahiha and S.~Banerjee, \emph{{A universal approach to Krylov State and
  Operator complexities}},  \href{https://arxiv.org/abs/2212.10583}{{\ttfamily
  2212.10583}}.

\bibitem{Avdoshkin:2022xuw}
A.~Avdoshkin, A.~Dymarsky and M.~Smolkin, \emph{{Krylov complexity in quantum
  field theory, and beyond}},
  \href{https://doi.org/10.1007/JHEP06(2024)066}{\emph{JHEP} {\bfseries 06}
  (2024) 066} [\href{https://arxiv.org/abs/2212.14429}{{\ttfamily
  2212.14429}}].

\bibitem{Camargo:2022rnt}
H.A.~Camargo, V.~Jahnke, K.-Y.~Kim and M.~Nishida, \emph{{Krylov complexity in
  free and interacting scalar field theories with bounded power spectrum}},
  \href{https://doi.org/10.1007/JHEP05(2023)226}{\emph{JHEP} {\bfseries 05}
  (2023) 226} [\href{https://arxiv.org/abs/2212.14702}{{\ttfamily
  2212.14702}}].

\bibitem{Bhattacharjee:2023uwx}
B.~Bhattacharjee, P.~Nandy and T.~Pathak, \emph{{Operator dynamics in
  Lindbladian SYK: a Krylov complexity perspective}},
  \href{https://doi.org/10.1007/JHEP01(2024)094}{\emph{JHEP} {\bfseries 01}
  (2024) 094} [\href{https://arxiv.org/abs/2311.00753}{{\ttfamily
  2311.00753}}].

\bibitem{Iizuka:2023pov}
N.~Iizuka and M.~Nishida, \emph{{Krylov complexity in the IP matrix model}},
  \href{https://doi.org/10.1007/JHEP11(2023)065}{\emph{JHEP} {\bfseries 11}
  (2023) 065} [\href{https://arxiv.org/abs/2306.04805}{{\ttfamily
  2306.04805}}].

\bibitem{Iizuka:2023fba}
N.~Iizuka and M.~Nishida, \emph{{Krylov complexity in the IP matrix model. Part
  II}}, \href{https://doi.org/10.1007/JHEP11(2023)096}{\emph{JHEP} {\bfseries
  11} (2023) 096} [\href{https://arxiv.org/abs/2308.07567}{{\ttfamily
  2308.07567}}].

\bibitem{Rabinovici:2023yex}
E.~Rabinovici, A.~S\'anchez-Garrido, R.~Shir and J.~Sonner, \emph{{A bulk
  manifestation of Krylov complexity}},
  \href{https://arxiv.org/abs/2305.04355}{{\ttfamily 2305.04355}}.

\bibitem{Zhang:2023wtr}
R.~Zhang and H.~Zhai, \emph{{Universal Hypothesis of Autocorrelation Function
  from Krylov Complexity}},  \href{https://arxiv.org/abs/2305.02356}{{\ttfamily
  2305.02356}}.

\bibitem{Hashimoto:2023swv}
K.~Hashimoto, K.~Murata, N.~Tanahashi and R.~Watanabe, \emph{{Krylov complexity
  and chaos in quantum mechanics}},
  \href{https://doi.org/10.1007/JHEP11(2023)040}{\emph{JHEP} {\bfseries 11}
  (2023) 040} [\href{https://arxiv.org/abs/2305.16669}{{\ttfamily
  2305.16669}}].

\bibitem{Erdmenger:2023wjg}
J.~Erdmenger, S.-K.~Jian and Z.-Y.~Xian, \emph{{Universal chaotic dynamics from
  Krylov space}}, \href{https://doi.org/10.1007/JHEP08(2023)176}{\emph{JHEP}
  {\bfseries 08} (2023) 176}
  [\href{https://arxiv.org/abs/2303.12151}{{\ttfamily 2303.12151}}].

\bibitem{Bhattacharyya:2023dhp}
A.~Bhattacharyya, D.~Ghosh and P.~Nandi, \emph{{Operator growth and Krylov
  complexity in Bose-Hubbard model}},
  \href{https://doi.org/10.1007/JHEP12(2023)112}{\emph{JHEP} {\bfseries 12}
  (2023) 112} [\href{https://arxiv.org/abs/2306.05542}{{\ttfamily
  2306.05542}}].

\bibitem{Alishahiha:2024rwm}
M.~Alishahiha and M.J.~Vasli, \emph{{Thermalization in Krylov basis}},
  \href{https://doi.org/10.1140/epjc/s10052-025-13757-2}{\emph{Eur. Phys. J. C}
  {\bfseries 85} (2025) 39} [\href{https://arxiv.org/abs/2403.06655}{{\ttfamily
  2403.06655}}].

\bibitem{Menzler:2024ifs}
H.G.~Menzler and R.~Jha, \emph{{Krylov delocalization/localization across
  ergodicity breaking}},
  \href{https://doi.org/10.1103/PhysRevB.110.125137}{\emph{Phys. Rev. B}
  {\bfseries 110} (2024) 125137}
  [\href{https://arxiv.org/abs/2403.14384}{{\ttfamily 2403.14384}}].

\bibitem{Loc:2024oen}
T.Q.~Loc, \emph{{Lanczos spectrum for random operator growth}},
  \href{https://arxiv.org/abs/2402.07980}{{\ttfamily 2402.07980}}.

\bibitem{Nandy:2024htc}
P.~Nandy, A.S.~Matsoukas-Roubeas, P.~Mart\'\i{}nez-Azcona, A.~Dymarsky and
  A.~del Campo, \emph{{Quantum Dynamics in Krylov Space: Methods and
  Applications}},  \href{https://arxiv.org/abs/2405.09628}{{\ttfamily
  2405.09628}}.

\bibitem{Sanchez-Garrido:2024pcy}
A.~S\'anchez-Garrido, \emph{{On Krylov Complexity}}, Ph.D. thesis, U. Geneva
  (main), 2024.
\newblock \href{https://arxiv.org/abs/2407.03866}{{\ttfamily 2407.03866}}.

\bibitem{Balasubramanian:2022tpr}
V.~Balasubramanian, P.~Caputa, J.M.~Magan and Q.~Wu, \emph{{Quantum chaos and
  the complexity of spread of states}},
  \href{https://doi.org/10.1103/PhysRevD.106.046007}{\emph{Phys. Rev. D}
  {\bfseries 106} (2022) 046007}
  [\href{https://arxiv.org/abs/2202.06957}{{\ttfamily 2202.06957}}].

\bibitem{Bhattacharjee:2022qjw}
B.~Bhattacharjee, S.~Sur and P.~Nandy, \emph{{Probing quantum scars and weak
  ergodicity breaking through quantum complexity}},
  \href{https://doi.org/10.1103/PhysRevB.106.205150}{\emph{Phys. Rev. B}
  {\bfseries 106} (2022) 205150}
  [\href{https://arxiv.org/abs/2208.05503}{{\ttfamily 2208.05503}}].

\bibitem{Nandy:2023brt}
S.~Nandy, B.~Mukherjee, A.~Bhattacharyya and A.~Banerjee, \emph{{Quantum state
  complexity meets many-body scars}},
  \href{https://doi.org/10.1088/1361-648X/ad1a7b}{\emph{J. Phys. Condens.
  Matter} {\bfseries 36} (2024) 155601}
  [\href{https://arxiv.org/abs/2305.13322}{{\ttfamily 2305.13322}}].

\bibitem{Banerjee:2022ime}
A.~Banerjee, A.~Bhattacharyya, P.~Drashni and S.~Pawar, \emph{{From CFTs to
  theories with Bondi-Metzner-Sachs symmetries: Complexity and
  out-of-time-ordered correlators}},
  \href{https://doi.org/10.1103/PhysRevD.106.126022}{\emph{Phys. Rev. D}
  {\bfseries 106} (2022) 126022}
  [\href{https://arxiv.org/abs/2205.15338}{{\ttfamily 2205.15338}}].

\bibitem{PhysRevD.108.025013}
A.~Chattopadhyay, A.~Mitra and H.J.R.~van Zyl, \emph{Spread complexity as
  classical dilaton solutions},
  \href{https://doi.org/10.1103/PhysRevD.108.025013}{\emph{Phys. Rev. D}
  {\bfseries 108} (2023) 025013}.

\bibitem{Nizami:2023dkf}
A.A.~Nizami and A.W.~Shrestha, \emph{{Krylov construction and complexity for
  driven quantum systems}},
  \href{https://doi.org/10.1103/PhysRevE.108.054222}{\emph{Phys. Rev. E}
  {\bfseries 108} (2023) 054222}
  [\href{https://arxiv.org/abs/2305.00256}{{\ttfamily 2305.00256}}].

\bibitem{PhysRevB.108.104311}
K.~Pal, K.~Pal, A.~Gill and T.~Sarkar, \emph{Time evolution of spread
  complexity and statistics of work done in quantum quenches},
  \href{https://doi.org/10.1103/PhysRevB.108.104311}{\emph{Phys. Rev. B}
  {\bfseries 108} (2023) 104311}.

\bibitem{PhysRevB.109.014312}
M.~Gautam, K.~Pal, K.~Pal, A.~Gill, N.~Jaiswal and T.~Sarkar, \emph{Spread
  complexity evolution in quenched interacting quantum systems},
  \href{https://doi.org/10.1103/PhysRevB.109.014312}{\emph{Phys. Rev. B}
  {\bfseries 109} (2024) 014312}.

\bibitem{PhysRevB.109.104303}
A.~Gill, K.~Pal, K.~Pal and T.~Sarkar, \emph{Complexity in two-point
  measurement schemes},
  \href{https://doi.org/10.1103/PhysRevB.109.104303}{\emph{Phys. Rev. B}
  {\bfseries 109} (2024) 104303}.

\bibitem{PhysRevLett.132.160402}
B.~Craps, O.~Evnin and G.~Pascuzzi, \emph{A relation between krylov and nielsen
  complexity},
  \href{https://doi.org/10.1103/PhysRevLett.132.160402}{\emph{Phys. Rev. Lett.}
  {\bfseries 132} (2024) 160402}.

\bibitem{Caputa:2024vrn}
P.~Caputa, H.-S.~Jeong, S.~Liu, J.F.~Pedraza and L.-C.~Qu, \emph{{Krylov
  complexity of density matrix operators}},
  \href{https://doi.org/10.1007/JHEP05(2024)337}{\emph{JHEP} {\bfseries 05}
  (2024) 337} [\href{https://arxiv.org/abs/2402.09522}{{\ttfamily
  2402.09522}}].

\bibitem{PhysRevB.110.064318}
B.~Zhou and S.~Chen, \emph{Spread complexity and dynamical transition in
  multimode bose-einstein condensates},
  \href{https://doi.org/10.1103/PhysRevB.110.064318}{\emph{Phys. Rev. B}
  {\bfseries 110} (2024) 064318}.

\bibitem{Nizami:2024ltk}
A.A.~Nizami and A.W.~Shrestha, \emph{{Spread complexity and quantum chaos for
  periodically driven spin chains}},
  \href{https://doi.org/10.1103/PhysRevE.110.034201}{\emph{Phys. Rev. E}
  {\bfseries 110} (2024) 034201}
  [\href{https://arxiv.org/abs/2405.16182}{{\ttfamily 2405.16182}}].

\bibitem{Balasubramanian:2023kwd}
V.~Balasubramanian, J.M.~Magan and Q.~Wu, \emph{{Quantum chaos, integrability,
  and late times in the Krylov basis}},
  \href{https://doi.org/10.1103/PhysRevE.111.014218}{\emph{Phys. Rev. E}
  {\bfseries 111} (2025) 014218}
  [\href{https://arxiv.org/abs/2312.03848}{{\ttfamily 2312.03848}}].

\bibitem{Baggioli:2024wbz}
M.~Baggioli, K.-B.~Huh, H.-S.~Jeong, K.-Y.~Kim and J.F.~Pedraza, \emph{{Krylov
  complexity as an order parameter for quantum chaotic-integrable
  transitions}},  \href{https://arxiv.org/abs/2407.17054}{{\ttfamily
  2407.17054}}.

\bibitem{Fu:2024fdm}
Y.~Fu, K.-Y.~Kim, K.~Pal and K.~Pal, \emph{{Statistics and Complexity of
  Wavefunction Spreading in Quantum Dynamical Systems}},
  \href{https://arxiv.org/abs/2411.09390}{{\ttfamily 2411.09390}}.

\bibitem{Caputa2022PRB}
P.~Caputa and S.~Liu, \emph{Quantum complexity and topological phases of
  matter}, \href{https://doi.org/10.1103/PhysRevB.106.195125}{\emph{Phys. Rev.
  B} {\bfseries 106} (2022) 195125}.

\bibitem{Anegawa:2024wov}
T.~Anegawa, N.~Iizuka and M.~Nishida, \emph{{Krylov complexity as an order
  parameter for deconfinement phase transitions at large N}},
  \href{https://doi.org/10.1007/JHEP04(2024)119}{\emph{JHEP} {\bfseries 04}
  (2024) 119} [\href{https://arxiv.org/abs/2401.04383}{{\ttfamily
  2401.04383}}].

\bibitem{Bhattacharya:2023zqt}
A.~Bhattacharya, P.~Nandy, P.P.~Nath and H.~Sahu, \emph{{On Krylov complexity
  in open systems: an approach via bi-Lanczos algorithm}},
  \href{https://doi.org/10.1007/JHEP12(2023)066}{\emph{JHEP} {\bfseries 12}
  (2023) 066} [\href{https://arxiv.org/abs/2303.04175}{{\ttfamily
  2303.04175}}].

\bibitem{Bhattacharyya:2023grv}
A.~Bhattacharyya, S.S.~Haque, G.~Jafari, J.~Murugan and D.~Rapotu,
  \emph{{Krylov complexity and spectral form factor for noisy random matrix
  models}}, \href{https://doi.org/10.1007/JHEP10(2023)157}{\emph{JHEP}
  {\bfseries 10} (2023) 157}
  [\href{https://arxiv.org/abs/2307.15495}{{\ttfamily 2307.15495}}].

\bibitem{Carolan:2024wov}
E.~Carolan, A.~Kiely, S.~Campbell and S.~Deffner, \emph{{Operator growth and
  spread complexity in open quantum systems}},
  \href{https://doi.org/10.1209/0295-5075/ad5b17}{\emph{EPL} {\bfseries 147}
  (2024) 38002} [\href{https://arxiv.org/abs/2404.03529}{{\ttfamily
  2404.03529}}].

\bibitem{Bhattacharya:2023yec}
A.~Bhattacharya, R.N.~Das, B.~Dey and J.~Erdmenger, \emph{{Spread complexity
  for measurement-induced non-unitary dynamics and Zeno effect}},
  \href{https://doi.org/10.1007/JHEP03(2024)179}{\emph{JHEP} {\bfseries 03}
  (2024) 179} [\href{https://arxiv.org/abs/2312.11635}{{\ttfamily
  2312.11635}}].

\bibitem{Bhattacharya:2024hto}
A.~Bhattacharya, R.N.~Das, B.~Dey and J.~Erdmenger, \emph{{Spread complexity
  and localization in PT-symmetric systems}},
  \href{https://doi.org/10.1103/PhysRevB.110.064320}{\emph{Phys. Rev. B}
  {\bfseries 110} (2024) 064320}
  [\href{https://arxiv.org/abs/2406.03524}{{\ttfamily 2406.03524}}].

\bibitem{Sahu:2024urf}
H.~Sahu, A.~Bhattacharya and P.P.~Nath, \emph{{Quantum complexity and
  localization in random quantum circuits}},
  \href{https://arxiv.org/abs/2409.03656}{{\ttfamily 2409.03656}}.

\bibitem{skinner2019measurementinduced}
B.~Skinner, J.~Ruhman and A.~Nahum, \emph{Measurement-induced phase transitions
  in the dynamics of entanglement},
  \href{https://doi.org/10.1103/PhysRevX.9.031009}{\emph{Phys. Rev. X}
  {\bfseries 9} (2019) 031009}.

\bibitem{li2018quantum}
Y.~Li, X.~Chen and M.P.A.~Fisher, \emph{Quantum zeno effect and the many-body
  entanglement transition},
  \href{https://doi.org/10.1103/PhysRevB.98.205136}{\emph{Phys. Rev. B}
  {\bfseries 98} (2018) 205136}.

\bibitem{li2019measurementdriven}
Y.~Li, X.~Chen and M.P.A.~Fisher, \emph{Measurement-driven entanglement
  transition in hybrid quantum circuits},
  \href{https://doi.org/10.1103/PhysRevB.100.134306}{\emph{Phys. Rev. B}
  {\bfseries 100} (2019) 134306}.

\bibitem{chan2019unitaryprojective}
A.~Chan, R.M.~Nandkishore, M.~Pretko and G.~Smith, \emph{Unitary-projective
  entanglement dynamics},
  \href{https://doi.org/10.1103/PhysRevB.99.224307}{\emph{Phys. Rev. B}
  {\bfseries 99} (2019) 224307}.

\bibitem{boorman2022diagnostics}
T.~Boorman, M.~Szyniszewski, H.~Schomerus and A.~Romito, \emph{Diagnostics of
  entanglement dynamics in noisy and disordered spin chains via the
  measurement-induced steady-state entanglement transition},
  \href{https://doi.org/10.1103/PhysRevB.105.144202}{\emph{Phys. Rev. B}
  {\bfseries 105} (2022) 144202}.

\bibitem{Biella2021manybodyquantumzeno}
A.~Biella and M.~Schir{\'{o}}, \emph{Many-{B}ody {Q}uantum {Z}eno {E}ffect and
  {M}easurement-{I}nduced {S}ubradiance {T}ransition},
  \href{https://doi.org/10.22331/q-2021-08-19-528}{\emph{{Quantum}} {\bfseries
  5} (2021) 528}.

\bibitem{szyniszewski2020universality}
M.~Szyniszewski, A.~Romito and H.~Schomerus, \emph{Universality of entanglement
  transitions from stroboscopic to continuous measurements},
  \href{https://doi.org/10.1103/PhysRevLett.125.210602}{\emph{Phys. Rev. Lett.}
  {\bfseries 125} (2020) 210602}.

\bibitem{barratt2022transitions}
F.~Barratt, U.~Agrawal, A.C.~Potter, S.~Gopalakrishnan and R.~Vasseur,
  \emph{Transitions in the learnability of global charges from local
  measurements},  2022.

\bibitem{zabalo2022infinite}
A.~Zabalo, J.H.~Wilson, M.J.~Gullans, R.~Vasseur, S.~Gopalakrishnan, D.A.~Huse
  et~al., \emph{Infinite-randomness criticality in monitored quantum dynamics
  with static disorder},  2022.

\bibitem{barratt2022field}
F.~Barratt, U.~Agrawal, S.~Gopalakrishnan, D.A.~Huse, R.~Vasseur and
  A.C.~Potter, \emph{Field theory of charge sharpening in symmetric monitored
  quantum circuits},
  \href{https://doi.org/10.1103/PhysRevLett.129.120604}{\emph{Phys. Rev. Lett.}
  {\bfseries 129} (2022) 120604}.

\bibitem{lunt2020measurement}
O.~Lunt and A.~Pal, \emph{Measurement-induced entanglement transitions in
  many-body localized systems},
  \href{https://doi.org/10.1103/PhysRevResearch.2.043072}{\emph{Phys. Rev.
  Research} {\bfseries 2} (2020) 043072}.

\bibitem{turkeshi2021measurementinduced}
X.~Turkeshi, \emph{Measurement-induced criticality as a data-structure
  transition},  2021.

\bibitem{zabalo2022operator}
A.~Zabalo, M.J.~Gullans, J.H.~Wilson, R.~Vasseur, A.W.W.~Ludwig,
  S.~Gopalakrishnan et~al., \emph{Operator scaling dimensions and
  multifractality at measurement-induced transitions},
  \href{https://doi.org/10.1103/PhysRevLett.128.050602}{\emph{Phys. Rev. Lett.}
  {\bfseries 128} (2022) 050602}.

\bibitem{iaconis2021multifractality}
J.~Iaconis and X.~Chen, \emph{Multifractality in nonunitary random dynamics},
  \href{https://doi.org/10.1103/PhysRevB.104.214307}{\emph{Phys. Rev. B}
  {\bfseries 104} (2021) 214307}.

\bibitem{Sierant2022dissipativefloquet}
P.~Sierant, G.~Chiriac{\`{o}}, F.M.~Surace, S.~Sharma, X.~Turkeshi, M.~Dalmonte
  et~al., \emph{Dissipative {F}loquet {D}ynamics: from {S}teady {S}tate to
  {M}easurement {I}nduced {C}riticality in {T}rapped-ion {C}hains},
  \href{https://doi.org/10.22331/q-2022-02-02-638}{\emph{{Quantum}} {\bfseries
  6} (2022) 638}.

\bibitem{bao2020theory}
Y.~Bao, S.~Choi and E.~Altman, \emph{Theory of the phase transition in random
  unitary circuits with measurements},
  \href{https://doi.org/10.1103/PhysRevB.101.104301}{\emph{Phys. Rev. B}
  {\bfseries 101} (2020) 104301}.

\bibitem{choi2020quantum}
S.~Choi, Y.~Bao, X.-L.~Qi and E.~Altman, \emph{Quantum error correction in
  scrambling dynamics and measurement-induced phase transition},
  \href{https://doi.org/10.1103/PhysRevLett.125.030505}{\emph{Phys. Rev. Lett.}
  {\bfseries 125} (2020) 030505}.

\bibitem{szyniszewski2019entanglement}
M.~Szyniszewski, A.~Romito and H.~Schomerus, \emph{Entanglement transition from
  variable-strength weak measurements},
  \href{https://doi.org/10.1103/PhysRevB.100.064204}{\emph{Phys. Rev. B}
  {\bfseries 100} (2019) 064204}.

\bibitem{block2022measurementinduced}
M.~Block, Y.~Bao, S.~Choi, E.~Altman and N.Y.~Yao, \emph{Measurement-induced
  transition in long-range interacting quantum circuits},
  \href{https://doi.org/10.1103/PhysRevLett.128.010604}{\emph{Phys. Rev. Lett.}
  {\bfseries 128} (2022) 010604}.

\bibitem{jian2020measurementinduced}
C.-M.~Jian, Y.-Z.~You, R.~Vasseur and A.W.W.~Ludwig, \emph{Measurement-induced
  criticality in random quantum circuits},
  \href{https://doi.org/10.1103/PhysRevB.101.104302}{\emph{Phys. Rev. B}
  {\bfseries 101} (2020) 104302}.

\bibitem{agrawal2021entanglement}
U.~Agrawal, A.~Zabalo, K.~Chen, J.H.~Wilson, A.C.~Potter, J.H.~Pixley et~al.,
  \emph{Entanglement and charge-sharpening transitions in u(1) symmetric
  monitored quantum circuits},
  \href{https://doi.org/10.1103/PhysRevX.12.041002}{\emph{Phys. Rev. X}
  {\bfseries 12} (2022) 041002}.

\bibitem{gullans2020scalable}
M.J.~Gullans and D.A.~Huse, \emph{Scalable probes of measurement-induced
  criticality},
  \href{https://doi.org/10.1103/PhysRevLett.125.070606}{\emph{Phys. Rev. Lett.}
  {\bfseries 125} (2020) 070606}.

\bibitem{sharma2022measurementinduced}
S.~Sharma, X.~Turkeshi, R.~Fazio and M.~Dalmonte, \emph{{Measurement-induced
  criticality in extended and long-range unitary circuits}},
  \href{https://doi.org/10.21468/SciPostPhysCore.5.2.023}{\emph{SciPost Phys.
  Core} {\bfseries 5} (2022) 023}.

\bibitem{zabalo2020critical}
A.~Zabalo, M.J.~Gullans, J.H.~Wilson, S.~Gopalakrishnan, D.A.~Huse and
  J.H.~Pixley, \emph{Critical properties of the measurement-induced transition
  in random quantum circuits},
  \href{https://doi.org/10.1103/PhysRevB.101.060301}{\emph{Phys. Rev. B}
  {\bfseries 101} (2020) 060301}.

\bibitem{vasseur2019entanglement}
R.~Vasseur, A.C.~Potter, Y.-Z.~You and A.W.W.~Ludwig, \emph{Entanglement
  transitions from holographic random tensor networks},
  \href{https://doi.org/10.1103/PhysRevB.100.134203}{\emph{Phys. Rev. B}
  {\bfseries 100} (2019) 134203}.

\bibitem{li2021conformal}
Y.~Li, X.~Chen, A.W.W.~Ludwig and M.P.A.~Fisher, \emph{Conformal invariance and
  quantum nonlocality in critical hybrid circuits},
  \href{https://doi.org/10.1103/PhysRevB.104.104305}{\emph{Phys. Rev. B}
  {\bfseries 104} (2021) 104305}.

\bibitem{turkeshi2020measurementinduced}
X.~Turkeshi, R.~Fazio and M.~Dalmonte, \emph{Measurement-induced criticality in
  $(2+1)$-dimensional hybrid quantum circuits},
  \href{https://doi.org/10.1103/PhysRevB.102.014315}{\emph{Phys. Rev. B}
  {\bfseries 102} (2020) 014315}.

\bibitem{lunt2021measurementinduced}
O.~Lunt, M.~Szyniszewski and A.~Pal, \emph{Measurement-induced criticality and
  entanglement clusters: A study of one-dimensional and two-dimensional
  clifford circuits},
  \href{https://doi.org/10.1103/PhysRevB.104.155111}{\emph{Phys. Rev. B}
  {\bfseries 104} (2021) 155111}.

\bibitem{sierant2022universal}
P.~Sierant and X.~Turkeshi, \emph{Universal behavior beyond multifractality of
  wave functions at measurement-induced phase transitions},
  \href{https://doi.org/10.1103/PhysRevLett.128.130605}{\emph{Phys. Rev. Lett.}
  {\bfseries 128} (2022) 130605}.

\bibitem{gullans2020dynamical}
M.J.~Gullans and D.A.~Huse, \emph{Dynamical purification phase transition
  induced by quantum measurements},
  \href{https://doi.org/10.1103/PhysRevX.10.041020}{\emph{Phys. Rev. X}
  {\bfseries 10} (2020) 041020}.

\bibitem{PhysRevB.107.L220201}
K.~Yamamoto and R.~Hamazaki, \emph{Localization properties in disordered
  quantum many-body dynamics under continuous measurement},
  \href{https://doi.org/10.1103/PhysRevB.107.L220201}{\emph{Phys. Rev. B}
  {\bfseries 107} (2023) L220201}.

\bibitem{Wiseman2009}
H.M.~Wiseman and G.J.~Milburn, \emph{Quantum Measurement and Control},
  Cambridge University Press (2009),
  \href{https://doi.org/https://doi.org/10.1017/CBO9780511813948}{https://doi.org/10.1017/CBO9780511813948}.

\bibitem{carmichael2009open}
H.~Carmichael, \emph{An open systems approach to quantum optics}, Springer,
  Berlin, Germany (1993),
  \href{https://doi.org/10.1007/978-3-540-47620-7}{10.1007/978-3-540-47620-7}.

\bibitem{gardiner2004quantum}
C.~Gardiner and P.~Zoller, \emph{Quantum noise}, Springer Science \& Business
  Media (2004).

\bibitem{daley2014quantum}
A.J.~Daley, \emph{Quantum trajectories and open many-body quantum systems},
  \href{https://doi.org/10.1080/00018732.2014.933502}{\emph{Adv. Phys.}
  {\bfseries 63} (2014) 77}.

\bibitem{Jacobs2014}
K.~Jacobs, \emph{Quantum Measurement Theory and its Applications}, Cambridge
  University Press (Aug., 2014),
  \href{https://doi.org/10.1017/cbo9781139179027}{10.1017/cbo9781139179027}.

\bibitem{PhysRevB.105.205125}
K.~Yamamoto, M.~Nakagawa, M.~Tezuka, M.~Ueda and N.~Kawakami, \emph{Universal
  properties of dissipative tomonaga-luttinger liquids: Case study of a
  non-hermitian xxz spin chain},
  \href{https://doi.org/10.1103/PhysRevB.105.205125}{\emph{Phys. Rev. B}
  {\bfseries 105} (2022) 205125}.

\bibitem{fisher2022quantum}
M.P.A.~Fisher, V.~Khemani, A.~Nahum and S.~Vijay, \emph{Random quantum
  circuits},  2022.

\bibitem{lunt2021quantum}
O.~Lunt, J.~Richter and A.~Pal, \emph{Quantum simulation using noisy unitary
  circuits and measurements},  2021.

\bibitem{potter2021entanglement}
A.C.~Potter and R.~Vasseur, \emph{Entanglement dynamics in hybrid quantum
  circuits},  in \emph{Quantum Science and Technology}, p.~211, Springer
  International Publishing (2022),
  \href{https://doi.org/10.1007/978-3-031-03998-0_9}{DOI}.

\bibitem{alberton2021entanglement}
O.~Alberton, M.~Buchhold and S.~Diehl, \emph{Entanglement transition in a
  monitored free-fermion chain: From extended criticality to area law},
  \href{https://doi.org/10.1103/PhysRevLett.126.170602}{\emph{Phys. Rev. Lett.}
  {\bfseries 126} (2021) 170602}.

\bibitem{buchhold2021effective}
M.~Buchhold, Y.~Minoguchi, A.~Altland and S.~Diehl, \emph{Effective theory for
  the measurement-induced phase transition of dirac fermions},
  \href{https://doi.org/10.1103/PhysRevX.11.041004}{\emph{Phys. Rev. X}
  {\bfseries 11} (2021) 041004}.

\bibitem{muller2022measurementinduced}
T.~M\"uller, S.~Diehl and M.~Buchhold, \emph{Measurement-induced dark state
  phase transitions in long-ranged fermion systems},
  \href{https://doi.org/10.1103/PhysRevLett.128.010605}{\emph{Phys. Rev. Lett.}
  {\bfseries 128} (2022) 010605}.

\bibitem{turkeshi2021entanglement}
X.~Turkeshi, M.~Dalmonte, R.~Fazio and M.~Schir\`o, \emph{Entanglement
  transitions from stochastic resetting of non-hermitian quasiparticles},
  \href{https://doi.org/10.1103/PhysRevB.105.L241114}{\emph{Phys. Rev. B}
  {\bfseries 105} (2022) L241114}.

\bibitem{turkeshi2021measurementinduced2}
X.~Turkeshi, A.~Biella, R.~Fazio, M.~Dalmonte and M.~Schir\'o,
  \emph{Measurement-induced entanglement transitions in the quantum ising
  chain: From infinite to zero clicks},
  \href{https://doi.org/10.1103/PhysRevB.103.224210}{\emph{Phys. Rev. B}
  {\bfseries 103} (2021) 224210}.

\bibitem{Kalsi_2022}
T.~Kalsi, A.~Romito and H.~Schomerus, \emph{Three-fold way of entanglement
  dynamics in monitored quantum circuits},
  \href{https://doi.org/https://doi.org/10.1088/1751-8121/ac71e8}{\emph{J.
  Phys. A Math. Theor.} {\bfseries 55} (2022) 264009}.

\bibitem{kells2021topological}
G.~Kells, D.~Meidan and A.~Romito, ``Topological transitions with continuously
  monitored free fermions.''

\bibitem{fleckenstein2022nonhermitian}
C.~Fleckenstein, A.~Zorzato, D.~Varjas, E.J.~Bergholtz, J.H.~Bardarson and
  A.~Tiwari, \emph{Non-hermitian topology in monitored quantum circuits},
  \href{https://doi.org/10.1103/PhysRevResearch.4.L032026}{\emph{Phys. Rev.
  Research} {\bfseries 4} (2022) L032026}.

\bibitem{Zhang2022universal}
P.~Zhang, C.~Liu, S.-K.~Jian and X.~Chen, \emph{Universal {E}ntanglement
  {T}ransitions of {F}ree {F}ermions with {L}ong-range {N}on-unitary
  {D}ynamics},
  \href{https://doi.org/10.22331/q-2022-05-27-723}{\emph{{Quantum}} {\bfseries
  6} (2022) 723}.

\bibitem{turk}
X.~Turkeshi, L.~Piroli and M.~Schir\'o, \emph{Enhanced entanglement negativity
  in boundary-driven monitored fermionic chains},
  \href{https://doi.org/10.1103/PhysRevB.106.024304}{\emph{Phys. Rev. B}
  {\bfseries 106} (2022) 024304}.

\bibitem{ashida2018full}
Y.~Ashida and M.~Ueda, \emph{Full-counting many-particle dynamics: Nonlocal and
  chiral propagation of correlations},
  \href{https://doi.org/10.1103/PhysRevLett.120.185301}{\emph{Phys. Rev. Lett.}
  {\bfseries 120} (2018) 185301}.

\bibitem{bacsi2021dynamics}
A.~B\'acsi and B.~D\'ora, \emph{Dynamics of entanglement after exceptional
  quantum quench},
  \href{https://doi.org/10.1103/PhysRevB.103.085137}{\emph{Phys. Rev. B}
  {\bfseries 103} (2021) 085137}.

\bibitem{dora2021correlations}
B.~D\'ora, D.~Sticlet and C.P.~Moca, \emph{Correlations at pt-symmetric quantum
  critical point},
  \href{https://doi.org/10.1103/PhysRevLett.128.146804}{\emph{Phys. Rev. Lett.}
  {\bfseries 128} (2022) 146804}.

\bibitem{gopalakrishnan2021entanglement}
S.~Gopalakrishnan and M.J.~Gullans, \emph{Entanglement and purification
  transitions in non-hermitian quantum mechanics},
  \href{https://doi.org/10.1103/PhysRevLett.126.170503}{\emph{Phys. Rev. Lett.}
  {\bfseries 126} (2021) 170503}.

\bibitem{10.21468/SciPostPhys.14.5.138}
Y.L.~Gal, X.~Turkeshi and M.~Schirò, \emph{{Volume-to-area law entanglement
  transition in a non-Hermitian free fermionic chain}},
  \href{https://doi.org/10.21468/SciPostPhys.14.5.138}{\emph{SciPost Phys.}
  {\bfseries 14} (2023) 138}.

\bibitem{Shi:2024bpu}
H.-L.~Shi, A.~Smerzi and L.~Pezz\`e, \emph{{Quantum Chaos, Randomness and
  Universal Scaling of Entanglement in Various Krylov Spaces}},
  \href{https://arxiv.org/abs/2407.11822}{{\ttfamily 2407.11822}}.

\bibitem{PhysRevLett.133.110201}
Y.~Chu, X.~Li and J.~Cai, \emph{Quantum delocalization on correlation
  landscape: The key to exponentially fast multipartite entanglement
  generation},
  \href{https://doi.org/10.1103/PhysRevLett.133.110201}{\emph{Phys. Rev. Lett.}
  {\bfseries 133} (2024) 110201}.

\bibitem{PhysRevB.103.085137}
A.~B\'acsi and B.~D\'ora, \emph{Dynamics of entanglement after exceptional
  quantum quench},
  \href{https://doi.org/10.1103/PhysRevB.103.085137}{\emph{Phys. Rev. B}
  {\bfseries 103} (2021) 085137}.

\bibitem{PhysRevB.97.045106}
S.~Lieu, \emph{Topological phases in the non-hermitian su-schrieffer-heeger
  model}, \href{https://doi.org/10.1103/PhysRevB.97.045106}{\emph{Phys. Rev. B}
  {\bfseries 97} (2018) 045106}.

\bibitem{PhysRevB.110.115135}
D.F.~Munoz-Arboleda, R.~Arouca and C.M.~Smith, \emph{Thermodynamics and
  entanglement entropy of the non-hermitian su-schrieffer-heeger model},
  \href{https://doi.org/10.1103/PhysRevB.110.115135}{\emph{Phys. Rev. B}
  {\bfseries 110} (2024) 115135}.

\bibitem{Rottoli_2024}
F.~Rottoli, M.~Fossati and P.~Calabrese, \emph{Entanglement hamiltonian in the
  non-hermitian ssh model},
  \href{https://doi.org/10.1088/1742-5468/ad4860}{\emph{Journal of Statistical
  Mechanics: Theory and Experiment} {\bfseries 2024} (2024) 063102}.

\bibitem{PhysRevLett.80.5243}
C.M.~Bender and S.~Boettcher, \emph{Real spectra in non-hermitian hamiltonians
  having {$\mathbf{P}\mathbf{T}$} symmetry},
  \href{https://doi.org/10.1103/PhysRevLett.80.5243}{\emph{Phys. Rev. Lett.}
  {\bfseries 80} (1998) 5243}.

\bibitem{viswanath1994recursion}
V.~Viswanath and G.~M{\"u}ller, \emph{The Recursion Method: Application to Many
  Body Dynamics}, Lecture Notes in Physics Monographs, Springer Berlin
  Heidelberg (1994).

\bibitem{bilanczos}
Y.~Saad, \emph{The lanczos biorthogonalization algorithm and other oblique
  projection methods for solving large unsymmetric systems}, {\emph{SIAM
  Journal on Numerical Analysis} {\bfseries 19} (1982) 485}.

\bibitem{PhysRevB.103.224210}
X.~Turkeshi, A.~Biella, R.~Fazio, M.~Dalmonte and M.~Schir\'o,
  \emph{Measurement-induced entanglement transitions in the quantum ising
  chain: From infinite to zero clicks},
  \href{https://doi.org/10.1103/PhysRevB.103.224210}{\emph{Phys. Rev. B}
  {\bfseries 103} (2021) 224210}.

\bibitem{10.21468/SciPostPhysCore.6.3.051}
C.~Zerba and A.~Silva, \emph{{Measurement phase transitions in the no-click
  limit as quantum phase transitions of a non-hermitean vacuum}},
  \href{https://doi.org/10.21468/SciPostPhysCore.6.3.051}{\emph{SciPost Phys.
  Core} {\bfseries 6} (2023) 051}.

\bibitem{Nguyen:2017yqw}
P.~Nguyen, T.~Devakul, M.G.~Halbasch, M.P.~Zaletel and B.~Swingle,
  \emph{{Entanglement of purification: from spin chains to holography}},
  \href{https://doi.org/10.1007/JHEP01(2018)098}{\emph{JHEP} {\bfseries 01}
  (2018) 098} [\href{https://arxiv.org/abs/1709.07424}{{\ttfamily
  1709.07424}}].

\bibitem{PhysRevLett.122.201601}
A.~Bhattacharyya, A.~Jahn, T.~Takayanagi and K.~Umemoto, \emph{Entanglement of
  purification in many body systems and symmetry breaking},
  \href{https://doi.org/10.1103/PhysRevLett.122.201601}{\emph{Phys. Rev. Lett.}
  {\bfseries 122} (2019) 201601}.

\bibitem{PhysRevResearch.3.013248}
H.A.~Camargo, L.~Hackl, M.P.~Heller, A.~Jahn, T.~Takayanagi and B.~Windt,
  \emph{Entanglement and complexity of purification in ($1+1$)-dimensional free
  conformal field theories},
  \href{https://doi.org/10.1103/PhysRevResearch.3.013248}{\emph{Phys. Rev.
  Res.} {\bfseries 3} (2021) 013248}.

\bibitem{PhysRevA.87.022310}
M.~Jiang, S.~Luo and S.~Fu, \emph{Channel-state duality},
  \href{https://doi.org/10.1103/PhysRevA.87.022310}{\emph{Phys. Rev. A}
  {\bfseries 87} (2013) 022310}.

\bibitem{Das:2024zuu}
R.N.~Das and T.~Mori, \emph{{Krylov complexity of purification}},
  \href{https://arxiv.org/abs/2408.00826}{{\ttfamily 2408.00826}}.

\bibitem{PhysRevB.109.224304}
P.H.S.~Bento, A.~del Campo and L.C.~C\'eleri, \emph{Krylov complexity and
  dynamical phase transition in the quenched lipkin-meshkov-glick model},
  \href{https://doi.org/10.1103/PhysRevB.109.224304}{\emph{Phys. Rev. B}
  {\bfseries 109} (2024) 224304}.

\bibitem{Wang_2014}
T.-L.~Wang, L.-N.~Wu, W.~Yang, G.-R.~Jin, N.~Lambert and F.~Nori, \emph{Quantum
  fisher information as a signature of the superradiant quantum phase
  transition}, \href{https://doi.org/10.1088/1367-2630/16/6/063039}{\emph{New
  Journal of Physics} {\bfseries 16} (2014) 063039}.

\bibitem{PhysRevB.100.184417}
S.~Yin, J.~Song, Y.~Zhang and S.~Liu, \emph{Quantum fisher information in
  quantum critical systems with topological characterization},
  \href{https://doi.org/10.1103/PhysRevB.100.184417}{\emph{Phys. Rev. B}
  {\bfseries 100} (2019) 184417}.

\bibitem{Poggi:2023mll}
P.M.~Poggi and M.H.~Mu\~noz Arias, \emph{{Measurement-induced
  multipartite-entanglement regimes in collective spin systems}},
  \href{https://doi.org/10.22331/q-2024-01-18-1229}{\emph{Quantum} {\bfseries
  8} (2024) 1229} [\href{https://arxiv.org/abs/2305.10209}{{\ttfamily
  2305.10209}}].

\bibitem{Pezze:2016nxl}
L.~Pezz\`e, A.~Smerzi, M.K.~Oberthaler, R.~Schmied and P.~Treutlein,
  \emph{{Quantum metrology with nonclassical states of atomic ensembles}},
  \href{https://doi.org/10.1103/RevModPhys.90.035005}{\emph{Rev. Mod. Phys.}
  {\bfseries 90} (2018) 035005}
  [\href{https://arxiv.org/abs/1609.01609}{{\ttfamily 1609.01609}}].

\bibitem{2014JPhA...47P4006T}
G.~{T{\'o}th} and I.~{Apellaniz}, \emph{{Quantum metrology from a quantum
  information science perspective}},
  \href{https://doi.org/10.1088/1751-8113/47/42/424006}{\emph{Journal of
  Physics A Mathematical General} {\bfseries 47} (2014) 424006}
  [\href{https://arxiv.org/abs/1405.4878}{{\ttfamily 1405.4878}}].

\bibitem{Chu:2023mzh}
Y.~Chu, X.~Li and J.~Cai, \emph{{Strong Quantum Metrological Limit from
  Many-Body Physics}},
  \href{https://doi.org/10.1103/PhysRevLett.130.170801}{\emph{Phys. Rev. Lett.}
  {\bfseries 130} (2023) 170801}
  [\href{https://arxiv.org/abs/2301.12113}{{\ttfamily 2301.12113}}].

\bibitem{Lira-Solanilla:2024alp}
A.~Lira-Solanilla, X.~Turkeshi and S.~Pappalardi, \emph{{Multipartite
  entanglement structure of monitored quantum circuits}},
  \href{https://arxiv.org/abs/2412.16062}{{\ttfamily 2412.16062}}.

\end{thebibliography}\endgroup

\end{document}